\documentclass[a4paper, 12pt]{article}
\pdfoutput=1
\usepackage{jheppub}

\usepackage[T1]{fontenc}
\usepackage{epsfig}
\usepackage{wrapfig}
\usepackage[caption=false, singlelinecheck=false, justification=raggedright]{subfig}
\usepackage{float}
\usepackage{array}
\usepackage{booktabs}
\usepackage{feynmp}
\usepackage{ulem}
\usepackage{soul,color}
\usepackage{amsmath}
\usepackage{amssymb}
\usepackage{amsfonts}
\usepackage{amsthm}
\usepackage{mathrsfs}
\usepackage{wasysym}
\usepackage{upgreek}
\usepackage{multirow}
\usepackage[table]{xcolor}
\usepackage{nicefrac}
\usepackage{extarrows}
\usepackage{txfonts}
\usepackage{paralist}
\usepackage{slashed}
\usepackage{tikz}
\usepackage{enumitem}
\usepackage{cancel}

\evensidemargin=\oddsidemargin
\newcommand{\be}{\begin{eqnarray*}} \newcommand{\ee}{\end{eqnarray*}}
\newcommand{\bea}{\begin{eqnarray}} \newcommand{\eea}{\end{eqnarray}}

\def\XXint#1#2#3{{\setbox0=\hbox{$#1{#2#3}{\int}$}
     \vcenter{\hbox{$#2#3$}}\kern-.5\wd0}}

\newcommand{\India}{The Institute of Mathematical Sciences (Retired), Taramani, Chennai 600113,India}
\newcommand{\Germany}{Institute for Theoretical Physics, University of Regensburg, Regensburg 93053, Germany}

\allowdisplaybreaks

\makeatletter
\gdef\@fpheader{}
\makeatother

\begin{document}

\title{Quantum Density of States and Integer Partitions: A Semiclassical Approach}

\author{M.~V.~N.~Murthy$^1$}
\affiliation[1]{\India}
\emailAdd{murthymundur@gmail.com}

\author{and Matthias Brack$^2$}
\affiliation[2]{\Germany}
\emailAdd{Matthias.Brack@physik.uni-regensburg.de}
%
\abstract{In this review we discuss semi-classical methods that are traditionally used to describe many-body systems in physics, but may also be used to describe partitions of integers in analytic number theory. Specifically, we explore the connection between the methods of statistical mechanics and number partitions.  Though the two fields appear very different, their fundamental issues bear a close resemblance. In the former case it is the distribution of a given amount of energy among the particles in an ensemble at a given temperature with well defined properties, while in the latter case it is the way an integer is partitioned into other integers, with or without restrictions. We begin with a discussion of the single-particle  quantum density of states, also called the level density, in which we illustrate the connection between the density of states and the classical periodic orbits through the semiclassical trace formula. This is then extended to many particle systems. We show that the asymptotic number partition is reproduced by the average (smooth) part of the level density at discrete integer values of the argument. In the especially interesting case of distinct square partitions, pronounced oscillations are well  reproduced by the periodic orbit theory in terms of a few orbits characterised by Pythagorean number triples. We speculate on the connection to Fermat's theorem as to why such regular oscillations (though vanishing asymptotically) exist only in this special case. Finally, we discuss some new results for integer partitions of primes, both unrestricted and distinct.}  
\vskip 1cm

\keywords{Semiclassical, Density of states, Partition function, Trace formula, Integer partitions}
\maketitle%

\newpage
\section{Introduction}
\label{sec:intro}

It was hundred years ago that the first bound-state problem, namely the Hydrogen atom, was solved using the newly minted quantum mechanics. Since then we have traversed a long way on the description of physical systems, using quantum mechanics as the fundamental dynamical framework. It is also slightly more than hundred years ago that the seminal paper of Hardy and Ramanujan \cite{bib:ramanujan} appeared on the problem of the asymptotic partition of integers. This asymptotic partition is formally identical to the density of states of a system obeying Bose statistics \cite{bib:bose} in a one-dimensional harmonic spectrum, which was formulated a few years later.  

It is therefore appropriate to look back and see what connects these apparently very different areas of research. We attempt to do this in a pedagogic review which is mostly based on published work done by the present authors together with other colleagues at different periods in different combinations. At the end, we shall also present some unpublished work on the distinct partitions of integers into primes.  

The first physical application of the Hardy-Ramanujan formula was done by Bohr and Kalcker \cite{bib:bohr-kalcker} who used it to estimate the density of energy levels of a heavy nucleus. The method that we follow here was first used by Bethe in 1936 \cite{bib:bethe}. In his seminal paper, he derived a formula for the nuclear level density using a statistical mechanics approach based on a gas of non-interacting fermions. This became the first model of the nucleus, known as the {\it Fermi gas model of the nucleus}. Bethe utilized the 'entropy' of the system and Sommerfeld's methods for calculating the properties of a degenerate gas. The formula for the level density that he derived bears a close resemblance to the Hardy-Ramanujan formula, as we show later. Bethe appears to have been unaware of the correspondence, since he did not refer to the work of Hardy and Ramanujan. The two approaches were shown to be mathematically equivalent in later work. 

The connection between the partition function in statistical mechanics and the generating function of partitions in number theory was made explicit by Auluck and Kothari \cite{bib:Auluck} in 1946. Following this, many other authors used the approach to obtain the Hardy-Ramanujan formula and to study other types of partitions, some ofthem using semiclassical methods \cite{bib:temperley, bib:julia, bib:nanda, bib:holthaus} leading to the study of physical properties of partition functions. We reproduce here a few examples of these results by combining them within a general formalism where the spectrum is more general and the occupancy of the levels going beyond Bose or Fermi statistics.

Our approach is to start by considering the simple case of a discrete quantum spectrum of a particle confined in an one-dimensional trap, and thereby to introduce some useful concepts. The spectrum is assumed to be integer-valued\footnote{This is easily done by an overall subtraction of a constant from the exact spectrum.}. We then introduce the many-particle quantum density of states. The problem of counting the number of excited states allowed at a given energy $E=n$, where $n$ is a positive integer, is identical to that of counting the number of partitions of an integer $E$ as a sum of its parts allowed by the spectrum. Consider, for example, the spectrum of an one-dimensional oscillator. In a many-particle system obeying Bose-Einstein statistics, we may consider the problem of distributing the particles in the oscillator levels at a given energy $E$. The number of ways the energy can be distributed is the same as the number of partitions $P_N(n)$, i.e., the number of ways of distributing $n=E$ as a sum of utmost $N$ integers, assuming that the energy of the lowest state is zero. Similarly, if the single-particle spectrum corresponds to that of a one-dimensional box, then we obtain the partition of $n$ as a sum of squares of integers. If the occupancy of each single-particle state is restricted to either 0 or 1 (instead of unrestricted distribution as in the bosonic case), we obtain distinct or fermionic partitions (with a {\it caveat} as explained later). In fact, the method may be extended to arbitrary occupancies such as in 'gentile statistics' \cite{bib:gentile}. 

The basic problem in quantum statistical mechanics is the partition of the energy, as a real number, among particles in an ensemble taking into account the nature of the single-particle spectrum. In this sense, the methods of statistical mechanics apply to all kinds of partitions of integers, each of them being a special case of the general problem in quantum statistical mechanics. While exact results are few, asymptotic properties of partitions may be obtained using semiclassical methods. Interestingly, while the relation between quantum statistical mechanics and partitions emerges naturally, classical statistical mechanics has no such analogy, unless a distinction is made in the ordering of integers in a partition, for example 3=1+2 being different from 3=2+1, which does not make sense for the partitioning of integers. In this sense, integer partition problem is truly a quantum phenomena.   

In this review we draw attention to semiclassical methods used in the context of partitions of an integer into sums involving other integers \cite{bib:muoi, bib:kutta, bib:bbbm, bib:mbbb, bib:mbb}.   In particular, we look at the special case of partitions of an integer into sums of distinct squares, which appears as a particular case since it displays regular oscillations with a prominent beat pattern even though there is no physical scale in the problem \cite{bib:mbbb}. Where are these oscillations coming from? The Fourier transform of the oscillating level density suggests a dominance of Pythagorean number triplets. As a consequence of Fermat's theorem, such triplets can only occur in square partitions. Hence the beat pattern is a particular feature of the distinct square partitions. This will be followed by a discussion of prime partitions, unrestricted as well as distinct, whereby we emphasise some new results published in \cite{bib:bbbm, bib:mbb}.

Section \ref{sec:spds} contains an introduction to the quantum level density of a single particle confined in a potential, and its relation to the partition function of the system. We discuss the semi-classical trace formula by which the level density is related to the classical primitive periodic orbits. In Sec. \ref{sec:mpds}, this is extended to the discussion of level the density of the many-particle system. We introduce a relation between the quantum statistical partition function and the many-particle density of states by means of its Laplace transform. Sections 2 and 3 are mainly based on readings from the book in reference \cite{bib:book}. We make a connection between integer partitions and the density of states when the energy is integer valued, and we derive a continuous trace formula for the level density using an integral representation of partitions. In Sec. \ref{sec:spcor}, we discuss some details of the general saddle-point method used to obtain the asymptotic density of states and present corrections to the smooth part of the level density, including higher orders depending on its convergence property. These results are then used in the later sections. 

Section \ref{sec:int-part} summarises the well-known asymptotic formulae for large $n$ of various partitions, following the work of Hardy and Ramanujan found in many text books \cite{bib:ayoub, bib:primetheorem,bib:eriksson}. The quantum level density is calculated for a general power-law spectrum, from which the well-known asymptotic results for the level density can be derived. We discuss in detail the cases of unrestricted (bosonic) and distinct (fermionic) partitions. The particles are confined to a potential whose spectrum is a monomial $m^s$. The general case in which each level supports utmost $k$ numbers of particles, the so-called 'gentile statistics' \cite{bib:gentile}, is treated in detail in  appendix \ref{appendix1}. We pay special attention to the case of distinct ($k=1$) square ($s=2$) partitions, where one finds pronounced oscillations with a beat-like structure when the discrete points at $n$ are joined by a continuous curve. None of the other known partitions display this behaviour. 

In Sect. \ref{sec:semiclass}, we present a detailed analysis of the oscillations in the distinct square partitions. We discuss a trace formula derived from the semiclassical periodic orbit theory (POT), which reproduces the oscillating part of the partitions, involving predominantly the Pythagorean number triplets that had been found in the Fourier transforms.

In section \ref{sec:prime}, we briefly present some new results on the partitions of a positive integer into sums of primes, both unrestricted and distinct. Some relevant details of the calculation are given appendix \ref{appendix2}. It is surprising that the prime partition has not received as much attention as it deserves \cite{bib:roth, bib:yang,bib:vaughan,bib:bbbm,bib:mbb}. We assume the particles to be moving in a potential whose quantum spectrum consists of only primes, prime gas as it were. There is no known Hamiltonian with a prime spectrum. It would be interesting to find a Hamiltonian with a confining potential whose quantum spectrum consists of primes. This poses a challenging problem for future research. 
 
We end the review with a summary in section \ref{sec:summary}.

\section{Single particle density of states}
\label{sec:spds}

Throughout this section we follow the notation and discussion as presented in the book \cite{bib:book}. We begin with some definitions and concepts using the single-particle density of states, also called the level density, which is defined as a sum over delta functions: 
\begin{equation}
g(E) := \sum_n \delta(E-E_n)\, .
\label{eq:gofE}
\end{equation}
Here $E_n$ are the energy eigenvalues of a particle moving in a potential $V({\bf r})$ obeying the Schrödinger equation
\begin{equation}
    \hat{H}\psi_n({\bf r})=[\hat{T}+V({\bf r})]\psi_n({\bf r})=E_n\psi_n({\bf r})\, .
    \label{eq:Schrodinger}
\end{equation}
Without loss of generality, we may set $E_n$ to be positive and integer valued by a constant shift. Depending on the symmetries of the Hamiltonian, the levels may be degenerate. The total number of levels up to energy $E$ is then given by
\begin{equation}
    N(E)=\int_0^E g(E')\, dE'\, .
\end{equation}
The level density $g(E)$ contains all the information about the nature of the quantum spectrum. In general, the density of states may be written as the sum of a smooth part and an oscillating part:
\begin{equation}
    g(E) = \tilde{g}(E)\,+\,\delta g(E) \, . 
    \label{eq:gsmoothgosc}
\end{equation}
This separation lends itself to a semi-classical analysis:
\begin{itemize}
    \item The smooth part $\tilde{g}(E)$ depends, in general, on the degrees of freedom of the system and the degeneracies which in turn depend on the symmetries of the Hamiltonian. It is usually determined semi-classically using the Extended Thomas-Fermi (ETF) model.
    \item The oscillating part $\delta g(E)$ may be calculated by a semi-classical "trace           formula" which reads
    \begin{equation}
    \delta g(E) = \sum_{\Gamma}\sum_{k=1}^\infty \mathcal{A}_{\Gamma k}\, 
    \cos\left[\frac{k}{\hbar}\,S_\Gamma(E) - \sigma_{\Gamma k}\, \frac{\pi}{2}\right].
    \label{eq:trace}    
    \end{equation}
    It relates the quantum spectrum of the Hamiltonian $\hat{H}$ to the properties of the periodic orbits of the corresponding classical Hamiltonian. Here $\Gamma$ counts the           classes of topologically distinct primitive periodic orbits (PPO), the index $k$  counts the number of repetitions of each PPO (or "harmonic"), 
          $S_\Gamma(E)= \oint \vec p_\Gamma.d\vec q_\Gamma$ is the classical action of the orbit $\Gamma$, and $\sigma_{\Gamma k}$ is the so-called Maslov index that determines the phase depending on the topology of $\Gamma$. The amplitude prefactor           $\mathcal{A}_{\Gamma k}$ depends on the energy, the time period and the stability of the orbit.   This correspondence between classical and quantum properties of the a system has often          been ignored. 
\end{itemize}
We will explore here this "classical-to-quantum correspondence" in specific cases without, however, going through detailed proofs \cite{bib:book}. 

The Laplace transform of the density of states $g(E)$ is by definition the single-particle canonical partition function $Z(\beta)$:
\begin{equation}
   Z(\beta) = \mathcal{L}_\beta[g(E)]=\int_0^\infty e^{-\beta E}\,g(E)\,dE = \sum_n e^{-\beta E_n}\, .
   \label{eq:zofb}
\end{equation}
In quantum statistics, $\beta=1/kT$ is proportional to an inverse temperature. Here it is simply a variable which has the dimension of an inverse energy. This connection between density of states and partition function is central to the discussions in the following sections. 

\subsection{Density of states for a general system}
Consider a system with an energy spectrum $E_n$ characterised by a single quantum number $n$. We may then write $E_n = f(n)$ and denote the degeneracy of each level by $D(n)$. The function $f(n)$ is assumed to be invertible such that $n = F(E_n)$. The density of states $g(E)$  in Eq.\ (\ref{eq:gofE}) can then be written as
\begin{equation}
    g(E)=\sum_{n_0}^\infty\delta(E-E_n)=D(E)\,|F'(E)|\,\sum_{n=0}^\infty \delta(n-F(E))\, ,
\end{equation}
since
\begin{equation}
\delta(E-E_n)=\delta(E-f(n))=|F'(E)|\,\delta(n-F(E))\, ,
\end{equation}
defining $D(E)=D(F(E))$.

Using the Poisson summation formula\footnote{The Poisson summation formula relates a smooth complex valued function $s(n)$ to its Fourier transform $S(k)$ through 
\begin{equation*}
    \sum_{n=-\infty}^\infty s(n) = \sum_{k=-\infty}^\infty S(k)
\end{equation*}} we may thus write the total level density as
\begin{equation}
    g(E)=\tilde{g}(E)+\delta g(E)=D(E)\,|F'(E)|\,\left[1+2\sum_{k=1}^\infty \cos[2\pi\,k\,F(E)]\right].
    \label{eq:gofegeneral}
\end{equation}
This is an important general formula for the level density applicable to all quantum systems with just one quantum number.  

Comparing the second (oscillating) part of Eq.\ (\ref{eq:gofegeneral}) with Eq.\ (\ref{eq:trace}), we conclude that $F(E)$ must be related to a classical action. This can be intuitively understood in the case of a (spatially) one-dimensional system whose states are non-degenerate. Indeed, the Bohr-Sommerfeld quantization condition in one dimension for the action is given by
\begin{equation}
    S(E_n)=\oint p(x)\,dx = 2\pi\hbar\,n = 2\pi \hbar\,F(E_n)
    \label{eq:clasact}
\end{equation}
and therefore 
\begin{equation}
    F(E)= \frac{1}{2\pi\hbar}\,S(E) \, .
\end{equation}
The oscillating part of the level density is thus given by
\begin{equation}
    \delta g(E) = 2\tilde{g}(E)\,\sum_{k=1}^\infty \cos\left[\frac{k}{\hbar}\,S(E)\right].
    \label{eq:trace2}
\end{equation}
The classical action $S(E)$ is calculated by integrating in (\ref{eq:clasact}) along the periodic orbit at the given energy $E$. The index $k$ denotes the number of revolutions around the orbit.  Note that this result holds for a simple one-dimensional system whose orbits are all periodic.  The amplitude factor $2\tilde{g}(E)$ depends on the derivative of the classical action which is related to the period of the orbit.

Let us demonstrate explicitly the connection between classical periodic orbits and the level density for the case of a one-dimensional harmonic oscillator. Its spectrum is given by 
\begin{equation}
    E_n/\hbar\omega = (n+1/2);   ~~~~~n=0,1,2,\cdots
    \label{eq:osc-spect}
\end{equation}
That is, we have here $F(E)=(E/\hbar\omega -1/2)$ and $F'(E)=1/\hbar\omega$. Consequently, the level density is given by
\begin{equation}
    g(E)=\frac{1}{\hbar\omega}\,\left[1+2\sum_{k=1}^\infty \cos\left[2\pi k 
\left(\frac{E}{\hbar\omega}-\frac{1}{2}\right)\right]\right]=\frac{1}{\hbar\omega}\,
\left[1+2\sum_{k=1}^\infty (-1)^k\,\cos\left(2\pi k\frac{E}{\hbar\omega}\right)\right] \, .
    \label{eq:1dho-tf}
\end{equation}
This is, in fact, an exact trace formula for the quantum level density of the one-dimensional harmonic oscillator as given in Ref.\cite{bib:book}, where many other exact trace formulae may also be found. 

The classical action may be obtained using the classical momentum $p(x)=\sqrt{2m[E-V(x)]}$ with the potential $V(x)=\frac{1}{2}\,m\omega^2\,x^2$. Substituting this into the action integral, we get
\begin{equation}
    S(E)=\oint p(x)\,dx = 2\int_{-x_0}^{x_0}p(x)\,dx=2\pi\frac{E}{\omega}\,,
\end{equation}
whereby the turning points are given by $\pm x_0=\pm \sqrt{2E/m}$. In the one-dimensional harmonic oscillator, there is only one class of periodic orbits. The sum is therefore over the primitive periodic orbit and its repetitions. For a general system, the sum extends over topologically distinct classes of primitive periodic orbits and their repetitions $k$, as expressed in Eq.\ (\ref{eq:trace}). 

We summarize: The density of states (or level density) of a quantum system, by definition, contains all the information about the quantum system. It can be expressed, at least to leading order, through a semi-classical trace formula as a sum over all periodic orbits of the corresponding classical system. This constitutes a classical-to-quantum correspondence.  For one-dimensional systems, the trace formula can become exact. 

The most chaotic classical systems have only isolated periodic orbits, if any at all. Surprisingly, it turns out that the quantum level density of such systems, at least to leading order, can be very simple \cite{bib:book,bib:gutz}, even when their classical dynamics is extremely complex and rich. 

\section{Density of states in a grand-canonical many-particle system}
\label{sec:mpds}

We now use the concepts and language of quantum statistics to establish the link between quantum physics and number theory.

Consider a system of $N$ {\it identical non-interacting particles} with well-defined exchange properties. We assume $N$ to be very large; in fact, we are thinking of the $N\rightarrow\infty$ limit. We shall use the term {\it many-particle system}, in contrast to the single-particle systems treated in the previous section, where $N$ can be an arbitrary (and even small) fixed number. The particles are assumed to be confined in a potential whose single-particle spectrum $\{\epsilon_n\}$ is assumed to be discrete and non-degenerate. Examples are a harmonic oscillator potential or an infinite square well in one dimension. At this stage of our presentation it is, however, not necessary to specify the exact form of the potential of the system. In fact, for our purpose it is enough to specify the single-particle spectrum without worrying about the structure of the Hamiltonian. We are interested in the density of states, denoted by $\rho(E)$, as a function of the total energy of the system. 

The total energy $E$ of the system is given by
\begin{equation}
    E= \sum_n d_n \epsilon_n \, .
\end{equation}
Here the sum runs over all allowed values of the single-particle energy $\epsilon_n$, and $d_n$ is the occupancy of the $n-$th level. If the particles are bosons, the occupancy $d_n$ may take any value including zero, where as for fermions it is either zero or one. In either case, the total number is given by 
\begin{equation}
    N = \sum_n d_n \, .  
\end{equation}
The total energy $E$ is thus obtained from distributions of $N$ particles of the spectrum $\{\epsilon_n\}$. There may be more than one way of distributing the total energy $E$ into a summand distributed over $N$ particles. If the energy $E$ is integer-valued, then this exercise is equivalent to the partitioning of integers as in number theory. This distribution of energy $E$ is in fact the main problem in quantum-statistical mechanics, where the partition function encodes all such information about the system. The canonical partition function $Z_N(\beta)$ of an $N$-body system is given by  
\begin{equation}
    Z_N(\beta)=\int_0^\infty dE\, \rho_N(E)\, \exp(-\beta E) \,,
    \label{eq:ZNbeta}
\end{equation}
where $\beta=1/kT$ is the inverse temperature in statistical thermodynamics and $\rho_N(E)$ is the many-particle density of states, the equivalent of $g(E)$ in the single-particle case, and $E$ is a dimensional real variable. In statistical mechanics, it represents the energy $E$. The density of states of a system of independent particles is denoted by $\rho(E)$, which is the equivalent of the single-particle density of states $g(E)$. For our general analysis here, we need not specify the form of $Z_N(\beta)$, as it depends on the system under consideration. 

For our purposes here we treat $\beta$ as a complex variable given by
\begin{equation}
\beta = x+i\,\tau, ~~~(x,\tau \, \in \, R) \, ,
\end{equation}
where $x$ and $\tau$ are real variables. We recognise that Eq.\ (\ref{eq:ZNbeta}) is formally the Laplace transform of the many-particle density of states $\rho_N(E)$ which may be obtained from a given partition function $Z_N(\beta)$ by its inverse Laplace transform: 
\begin{equation}
    \rho_N(E) =\frac{1}{2\pi\,i}\int_C d\beta\, Z_N(\beta)\, \exp(\beta E).
    \label{eq:rhoNE}
\end{equation}
The contour $C$ runs parallel to the imaginary axis $\tau$ with real part $x=\epsilon>0$, so that we may write
\begin{equation}
    \rho_N(E) =\frac{1}{2\pi}\int_{-\infty}^\infty d\tau\, Z_N(\epsilon+i\tau)\, e^{(\epsilon+i\tau) E}.
    \label{eq:rhoNE1}
\end{equation}
Formally Eqs. (\ref{eq:ZNbeta},\,\ref{eq:rhoNE}) have the same relation both in the single-particle  and the many-particle case.   

We note that, in the context of number theory, any given partition of an integer $n$ denoted as $P(n)$ is defined through the generating function
\begin{equation}
    Z(\beta) := \sum_{n=0}^\infty P(n)\, x^n\, ; ~ 0 < x < 1 \, .
    \label{eq:partgen}
\end{equation}
Identifying $x=e^{-\beta}$, this has the same form as the partition function in quantum statistics except that the continuous energy $E$ is replaced by an integer $n$. The form of Eqs. (\ref{eq:ZNbeta},\ref{eq:partgen}) immediately suggests that the partition $P(n)$ must be related to the quantum many-particle density $\rho(E)$ when $E=n$. In fact, using the above form of $Z(\beta)$ it is straightforward to find 
\begin{equation}
\rho(E)=\sum_{n=0}^\infty P(n) \, \delta(E-n) \, ,
\label{eq:partdens}
\end{equation}
where $\delta(E-n)$ is the Dirac delta function peaked at $E=n$. In order to recover the $P(n)$ from the density of states, we integrate over a small interval around $E=n$: 
\begin{equation}
    P(n)=\int_{n-a}^{n+a} \rho(E)  \, dE \,  ,~~~~~~(0<a<1)\, .  
\end{equation}

We explore further this relation between the quantum density of states and the partition of an integer $n$. We start with the expression for the quantum level density given in Eq.\ (\ref{eq:rhoNE}). We drop the superscript $N$ hereafter, since in the case of partitions we are interested in the limit $N\rightarrow\infty$. That is, we have an unlimited number of particles in the ensemble to choose from to divide the integer $n$ into its parts. The only restriction comes from the value of the integer $n$ itself and any condition on the occupancy. For example, consider the case of $n$ expressed as a sum of other integers. The largest number of parts is $n$ itself when it is expressed as $n=1+1+1+\cdots+1$ repeated $n$ times, when there is no restriction on the occupancy of the level (e.g. for bosons). 

We assume that the partition function $Z(\beta)$ has no poles on the right half of $\beta$ plane including the imaginary axis. We may therefore shift the contour C onto the imaginary axis, that is choose $\epsilon=0$ in Eq.\ (\ref{eq:rhoNE1}),
\begin{equation}
      \rho(E) =\frac{1}{2\pi}\int_{-\infty}^\infty d\tau\, Z(i\tau)\, e^{(iE\tau)}.
    \label{eq:rhoE}
\end{equation}
We now state the following result:
\paragraph{Lemma:} {\it The Laplace inversion integral in (\ref{eq:rhoE}) limited to the interval $t\in (-k\pi,+k\pi)$ with $k=1,2,\cdots$ yields a sum of Bessel functions weighted by $P(n)$ such that its value at $E=n$ is $k\,P(n)$}:
\begin{equation}
    \rho^{(k)}(E)=\frac{1}{2\pi} \int_{-k\pi}^{+k\pi} e^{iE\tau}Z(i\tau) \, dt 
    \,= \, k\sum_{m=0}^\infty \,P(m)\, j_0[k\pi(E-m)]\, ,
    \label{eq:rhokofE}
\end{equation}
where $j_0(x)=\sin(x)/x $ is the spherical Bessel function of order zero.

\noindent{\bf Proof:} For finite $k$, using (\ref{eq:partgen}) for $Z(\beta)$ and integrating over $\tau$ yields the above result Eq.\ (\ref{eq:rhokofE}).  Using the property of the Bessel function, $j_0(0)=1$, we obtain
\begin{equation}
\rho^{(k)}(E=n)\,= \, k P(n)\, ,
\end{equation}
which completes the proof. 

\begin{figure}[h]
    \centering
    \includegraphics[width=0.7\linewidth]{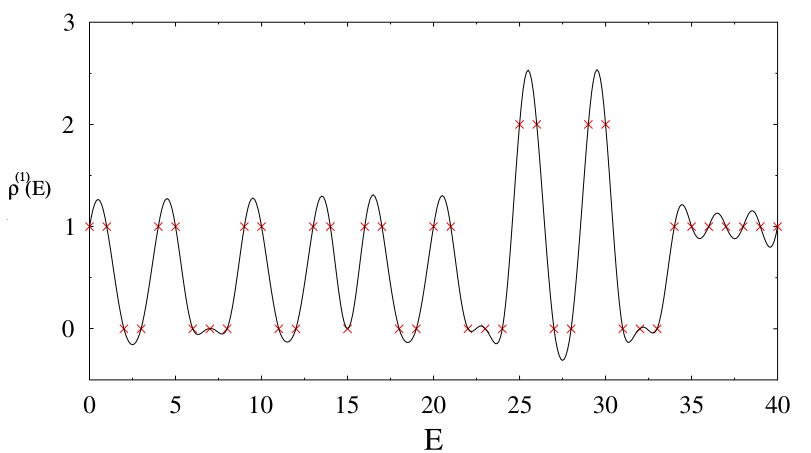}
    \caption{An example of the Bessel-smoothed partition density $\rho^{(1)}(E)$. The crosses(red) show the exact values of partitions $P(n)$ when $E=n$ expressed as a sum of  distinct squares (see section (\ref{sec:int-part}) for more details).}
    \label{fig:contpofn}
\end{figure}

It is easy to verify that in the limit $k\rightarrow \infty$, the Bessel function goes over to the delta function and we have
\begin{equation}
    \lim_{k\rightarrow\infty}\,[k\, j_0(k\pi(E-n))] = \delta(E-n)\, . 
\end{equation}
Therefore we have in this limit
\begin{equation}
\lim_{k\rightarrow\infty}\, \rho^{(k)}(E) \, =\sum_{n=0}^\infty \,P(n)\, \delta(E-n)\, = \rho(E) . 
\end{equation}
Note that the function $\rho^{(1)}(E)$ is a smooth function which interpolates the exact values of $P(n)$ whenever $E=n$.  We call this the {\it Bessel-smoothed} partition density.  It is now straightforward to obtain an {\it integral representation} of the partition $P(n)$:
\begin{equation}
   P(n) \,=\,\rho^{(1)}(n) \,=\, \frac{1}{2\pi} \int_{-\pi}^{+\pi} e^{iE\tau}Z(i\tau) \, d\tau   \label{eq:pofn}
\end{equation}
This is {\it an exact result}, as obtained in Ref.\ \cite{bib:mbbb} in a special case, and has not been  stated before in this form as far as we know. It provides a fast way of computing partitions of various kinds that can be performed numerically even on a laptop. An illustration of this is shown in Fig.(\ref{fig:contpofn}) for the special case of partition of integer $n$ as a sum of distinct squares.  

We conclude this section with a continuous trace formula for $\rho(E)$. In Eq.\ (\ref{eq:rhoE}) the imaginary part of the integrand is antisymmetric with respect to $\tau=0$. Therefore $\rho(E)$ is real, as it should be, and we need to retain only the real part of the integral. We get
\begin{equation}
    \rho(E)=\frac{1}{2\pi}\int_{-\infty}^{\infty} |Z(i\tau)|\, \cos[E\tau+\phi(\tau)]\, d\tau \, ,
\label{eq:rhotrace}
\end{equation}
where the phase $\phi(\tau)$ is given by
\begin{equation}
    \phi(\tau) = \mathcal{I}m[\ln\, Z(i\tau)]\, . 
\end{equation}
In its structure, the formula (\ref{eq:rhotrace}) resembles a semiclassical trace formula as given in Eq.\\ (\ref{eq:gofegeneral}). However, here we have a continuous integral instead of a discrete sum over periodic orbits. Hence we call it a {\it continuous trace formula} for the many-particle level density $\rho(E)$. Note that it is {\it exact}, although its practical use is not clear yet at this point. In Sect.\ \ref{sec:semiclass}, we shall start from Eq.\ (\ref{eq:rhotrace}) and show that using a series of approximations, a practically usable trace formula can be derived which reproduces the regular oscillations observed in distinct square partitions. For the one-dimensional harmonic oscillator, Eq.\ (\ref{eq:rhotrace}) reduces to the exact trace formula Eq.\ (\ref{eq:1dho-tf}).

In the next section we discuss a general formalism for obtaining an asymptotic level density, or equivalently, an asymptotic number partition.  We emphasise that the information about a specific type of partition is dictated by the choice of $Z(\beta)$ which depends on the specific system.

\section{Asymptotic expansion of the smooth part of the density}
\label{sec:spcor}

We present here a general formalism for obtaining the leading asymptotic form of $P(n)$ through an asymptotic expansion of $\rho(E)$ and show how next-to-leading order corrections can help to obtain a faster convergence to the average values of $P(n)$. We follow closely the method used in Ref.\ \cite{bib:muoi} with improvements given in \cite{bib:mbbb}. Since the average of $\rho(E=n)$ can be identified with that of $P(n)$ when the energy is an integer, the asymptotic analysis holds for both functions. 

Consider the integral in Eq.\ (\ref{eq:rhoE}). We write it in the following form 
\begin{equation}
\rho(E) =\frac{1}{2\pi i}\int_{\epsilon-i\infty}^{\epsilon+i\infty} {\rm d} \beta\,
 e^{S(E,\beta)} \, ,
\label{eq:asymsp1}
\end{equation}
where $S(E,\beta)$, in statistical mechanics called the {\it entropy function}, is given by
\begin{equation}
    S(E,\beta)\, = \, E\beta \,+\, \ln\, Z(\beta) \, .
    \label{eq:entropy}
\end{equation}
The asymptotic smooth part of the level density $\rho(E)$ is now obtained by performing a so-called "saddle-point integration". In most cases exact integration is not possible. The main contribution to the smooth part of the level density for large $E$ comes from the neighbourhood of a real saddle point, if it exists, in the complex $\beta$ plane, lying near the imaginary axis. Doing the integral over such a saddle point by the method of stationary-phase approximation yields the asymptotic smooth part of $\rho(E)$ and hence that of $P(n)$ (with $E=n$). In this section we briefly review the method of stationary-phase integration for a real saddle, including saddle-point corrections of higher order, closely following the method given in Ref.\ \cite{bib:jelovic}.

We denote the asymptotic form of the level density, valid for large $E$, by $\overline{\rho}(E)$. We first expand the exponent $S(E,\beta)$ around the saddle point (SP). For convenience, we denote the derivatives of $S$ here by
\begin{equation}
S_n(\beta)=\frac{\partial^n S(E,\beta)}{\partial \beta^n}\,,
\label{eq:asymsp3}
\end{equation} 
omitting the argument $E$ which for the present development is just a parameter.

We assume that there exists a real SP $\beta_0$ such that
\begin{equation}
S_1(\beta_0) = \frac{\partial S(E,\beta)}{\partial \beta} = 0\,, \qquad \beta_0 > 0\,.
\end{equation}
Taylor expanding the entropy $S(E,\beta)$ about $\beta_0$, we have
\begin{equation}
S(E,\beta) =S(E,\beta_0)+S_2(\beta_0)\left[\frac{(\beta-\beta_0)^2}{2}\right]+R(\beta)\,,
\label{eq:asymsp4}
\end{equation}
where
\begin{equation}
R(\beta)=\sum_{m=3}^{\infty}S_m(\beta_0)\left[\frac{(\beta-\beta_0)^m}{m!}\right]\,.
\label{eq:asymsp5}
\end{equation}
The asymptotic level density is then given by
\begin{equation}
\overline{\rho}(E) = \frac{e^{S(E,\beta_0)}}{2\pi i} \int_{\epsilon-i\infty}^{\epsilon+i\infty} 
{\rm d}\beta\,  e^{S_2(\beta_0)\left[\frac{(\beta-\beta_0)^2}{2!}\right]}\,e^{R(\beta)} \, .
\label{eq:asymsp6}
\end{equation}
Expanding the exponential of $R$ under the integral, we find
\begin{equation}
\overline{\rho}(E) = \frac{e^{S(\beta_0)}}{2\pi} \int_{-\infty}^{\infty} {\rm d}u\,  
           e^{-\frac12 S_2(\beta_0)\,u^2}
           \!\left[1+R(u)+\frac{R^2(u)}{2!}+\frac{R^3(u)}{3!}+\cdots\right] \, .
\label{eq:asymsp8}
\end{equation}
where $u=(\beta-\beta_0)/i\, .$ Only Gaussian integrals over even powers of $u$ survive. Collecting the even powers of $u$, we get
\begin{eqnarray}
\overline{\rho}(E) = \frac{e^{S(\beta_0)}}{2\pi} \int_{-\infty}^{\infty} {\rm d}u\,  
       e^{-\frac12 S_2(\beta_0)\, u^2}
       \!\left[1+\sum_{m=2}^\infty (-1)^m u^{2m}\sum_{\{k\}}
       \prod_{i=1}^k\left(\frac{S_{n_i}}{n_i!}\right)^{m_i}\!
       \left(\frac{1}{m_i!}\right)\right]\, ,
 \label{eq:asymsp9}
\end{eqnarray}
where the sum over $\{k\}$ involves the following constraint:
\begin{equation}
2m=m_1n_1+m_2n_2+\cdots + m_kn_k\,, \qquad n_i\ge 3\,, \qquad k\ge 1\,,
\label{eq:asymsp10}
\end{equation}
which is precisely the allowed number of partitions of $2m$ into $k$ parts, with repetitions allowed through the power $m_i$. All such partitions contribute at order $2m$ in $u$. The integration is now straightforward, and we obtain a result that formally contains SP contributions to all orders:
\begin{eqnarray}
\overline{\rho}(E) &=& \frac{e^{S(\beta_0)}}{\sqrt{2\pi S_2(\beta_0)}}\,
       \left[1+\sum_{m=2}^\infty (-1)^m \frac{(2m-1)!!}{ S_2^m}\sum_{\{k\}}
       \prod_{i=1}^k\left(\frac{S_{n_i}}{n_i!}\right)^{m_i}
       \left(\frac{1}{m_i!}\right)\right]\!. \nonumber\\
       &=&\frac{e^{S(\beta_0)}}{\sqrt{2\pi S_2(\beta_0)}}\,
       \left[1+\frac{3!!(S_4/4!)}{S_2^2} -\frac{5!!(S_3/3!)^2}{2!(S_2)^3}+\cdots\right]
\label{eq:asymsp11}
\end{eqnarray}
This is a series with $(1/S_2)$ as an expansion parameter. Even though this result is written to all orders, we note that it is an asymptotic series obtained by Taylor expansion of the entropy around a point $\beta= \beta_0$. The truncation of the series requires some care depending on the convergence of the series.  


\section{Partitions and quantum level density}
\label{sec:int-part}

Historically, one of the earliest applications of the quantum level-density formalism was by Bethe in 1936 \cite{bib:bethe}, who derived a formula for the nuclear level density which bears close resemblance to the Hardy-Ramanujan formula for integer partitions. The nuclear level-density formula estimates the number of available states for a nucleus at a given excitation energy, equivalent to the problem of distributing a given excitation energy  among particles without any restriction. Bethe's derivation used, for the first time, the Fermi gas model of the nucleus. He applied the methods of statistical mechanics relevant for a degenerate Fermi gas and used the saddle-point approximation to derive the following asymptotic formula for the level density:
\begin{equation}
\rho_{nucl}\ (E)\approx \frac{\sqrt\pi}{12({a})^{1/4}E^{5/4}}\exp(2\sqrt{aE})\,,
\label{eq:bethe-nucleus}
\end{equation}
where $a$ is the level-density parameter which depends on the nucleus, typically determined by the mean density of single particle levels at the Fermi energy. This formula is critical for understanding nuclear reactions, such as the capture of slow neutrons. The exponential increase seen above is the same as for the asymptotic density of states of bosons, while the prefactor has a different energy dependence. The following discussion broadly follows, in spirit, the method used by Bethe and others \cite{bib:Auluck, bib:temperley, bib:nanda, bib:holthaus} while expanding the scope to more general cases, concerning both spectra and occupancies \cite{bib:muoi}.

\begin{figure}[ht]
    \centering
    \includegraphics[width=0.7\linewidth]{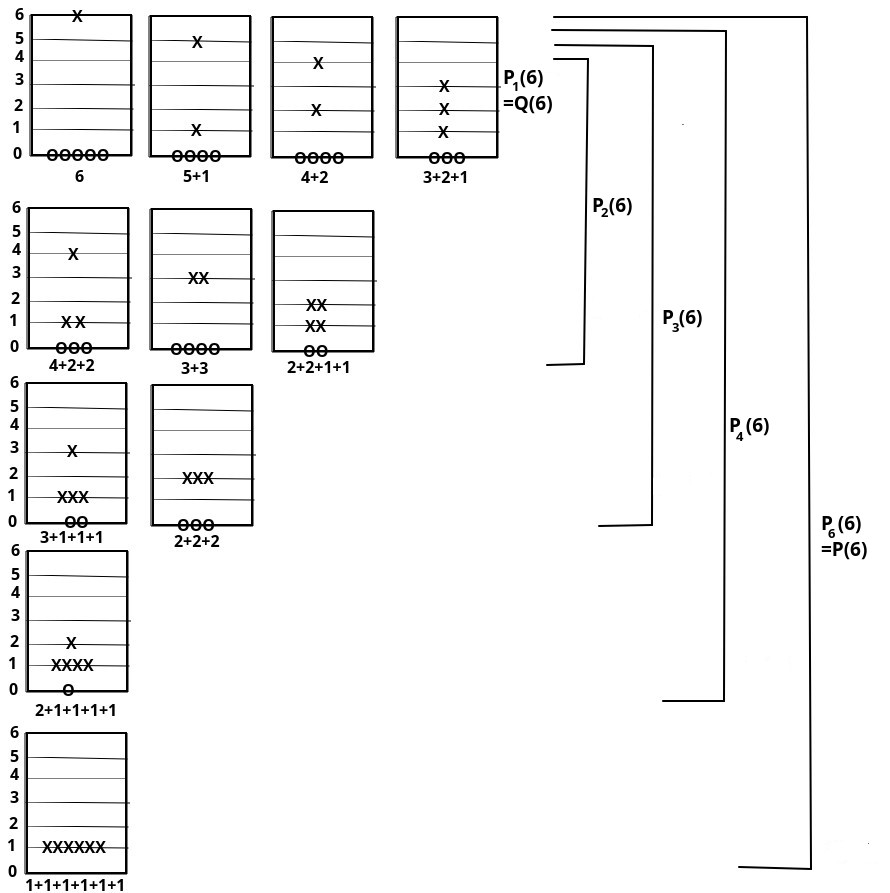}
    \caption{Identical bosons in an oscillator potential. Number of possibilities with excitation 
             energy $E=6$.}
    \label{fig:partbosons-6}
\end{figure}

Consider a system of identical bosons moving in a confining potential in one spatial dimension.  The simplest example of such a system is the one-dimensional harmonic oscillator. The spectrum of the 1d oscillator is given by Eq.\ (\ref{eq:osc-spect}). It is essentially an integer spectrum with $\epsilon_n=n$ with $n=0,1,2,\cdots$, expressed in dimensionless units ($\hbar=1=\omega$), after subtracting the zero-point energy of $1/2$. We define the ground state as one in which all particles are in the lowest zero-energy state. Let us now supply an excitation energy $E$ to the system. Note that this has to be an integer since the energy is quantised. This excitation energy can be distributed in many different ways: for example all the energy may be absorbed by a single particle to be excited to the state with energy $E$. On the other hand, all particles may gain unit energy and be placed in the first excited state. Many possibilities exist, as shown for the example where the system is given $6$ energy units in Fig.\ \ref{fig:partbosons-6}.
It is clear from that figure that the number of ways of distributing the energy among bosons is simply the number of ways of partitioning the energy ($E=n$) into a sum of smaller integers, i.e., $P(6)=11$. 

This is an unrestricted case in the sense that there is no constraint on the number of particles in a given energy level, except for the constraint imposed by the excited energy itself.  We call this an {\it unrestricted partition of an integer} $P(n)$. In the particular case of the harmonic oscillator, it grows exponentially, thus becoming very large soon. For example, $P(100)>10^8$. Hardy and Ramanujan\cite{bib:ramanujan} derived the now famous asymptotic expression for the growth of unrestricted integer partitions. Later, an exact formula for $P(n)$ in the form of a series was derived by Rademacher\cite{bib:rademacher}. 

We also notice from Fig.\ \ref{fig:partbosons-6} that we can make some distinctions within the set of unrestricted partitions. For example 
\begin{eqnarray}
\mbox{Distinct partitions}:~6,~5+1,~4+2,~3+2+1 &\Rightarrow& ~P_1(6)=Q(6)=4 \nonumber\\
\mbox{Up to 2 repetitions}:~ 3+3,~4+1+1,~2+2+1+1 &\Rightarrow& ~P_2(6)=7\nonumber\\
\mbox{Up to 3 repetitions}:~3+1+1+1,~2+2+2 &\Rightarrow&~P_3(6)=9 \nonumber\\
\mbox{Up to 4 repetitions}:~2+1+1+1+1 &\Rightarrow&~P_4(6)=10 \nonumber\\
\mbox{Up to 6 repetitions}:~1+1+1+1+1+1 &\Rightarrow&~P_6(6)=11=P(6)
\label{eq:part-6}
\end{eqnarray}
In general, $P_k(n)$ denotes the number of partitions of $n$ in each partition where an integer occurs utmost $k$ times. When $k=1$, an integer occurs exactly once in each summand of $n$. We call this a {\it distinct partition} of $n$ and denote it by $Q(n)$. The maximum number of unrestricted partitions allowed for any $n$ becomes simply $P(n)$. This is precisely the number of ways to distribute identical bosons at an excitation energy $E=n$. In section \ref{sec:mpds} above, it has been denoted as $\rho(E)$, the many particle density.  The one-dimensional harmonic trap is one of the many examples that illustrate the link between quantum-statistical mechanics and number theory. This link has been recognised for a long time \cite{bib:Auluck,bib:temperley,bib:nanda,bib:holthaus}\footnote{There exist other connections between physics and multiplicative number theory. Julia \cite{bib:julia} considered the so-called Riemann gas, whose partition function is the Riemann zeta function. Some of these concepts figure also in one-dimensional spin chains \cite{bib:knauf}}.  

Furthermore, we notice that the distinct partition $Q(n)$ bears a resemblance to the way identical fermions are distributed at a given excitation energy, as seen in top line of Fig.\ \ref{fig:partbosons-6}. Distinct partitions are thus obtained by restricting each integer in the summand to appear exactly once. This is equivalent to applying the Pauli principle to the distribution of particles sharing a given excitation energy, but with a {\it caveat}. Notice that, while the particle distribution in the single-particle energy levels follows the Pauli principle, the ground state does not reflect the distribution in the Fermi sea if the particles are truly fermions. This is unphysical if treated as a Fermi gas, but it leads to the distinct partitions of the excitation energy. Clearly, to make a connection to the distinct number partition, we should imagine that the particle space is occupied obeying the Pauli principle and collapse the states below the Fermi energy into a single ground state, ignoring the Pauli principle. Physically, there is no system obeying such a hybrid rule. It is just a device that allows us to make the formal connection from statistical mechanics to distinct partitions. 

A general description of the quantum statistics of a system resulting in partitions of the type $P_k(n)$ with occupancy factor $k$ was, indeed, considered by Gentile \cite{bib:gentile} already in 1939. Gentile considered a system where the maximum occupancy of an energy level lies between fermionic (restricted to 1) and bosonic (unrestricted). He obtained the partition function of a gas of particles obeying the statistics determined by the occupancy $\le k$; see Eq.\ (\ref{eq:zprod}) in appendix\ (\ref{appendix1}), where we give the general description of partitions for arbitrary occupancies following Gentile statistics \cite{bib:kutta}. The partition function of Gentile statistics has the property that it nicely interpolates between Bose and Fermi statistics. The asymptotic formula obtained with Gentile statistics reduces to that for distinct partitions in the fermionic ($k=1$) limit, while the bosonic or unrestricted partitions are obtained by removing any restriction on the occupancy $k$. We shall in the following treat the two realistic cases, namely bosons ($k\rightarrow \infty$) and fermions ($k=1$). 

\subsection{Unrestricted integer partitions}

Consider ideal bosons confined in a potential with single-particle energies $\epsilon_m=m^s$, where $m=0,1,2,3,\ldots$ and the real power $s$ is kept arbitrary. The energy is measured in dimensionless units. In statistical mechanics, $s$ can be an arbitrary positive real number. In the context of number partitions, we assume $s$ to be a positive integer, $s=1,2,3,\cdots$. For example, for $s=1$ the spectrum can be mapped onto that of a simple harmonic oscillator system with $\hbar\omega=1$. For $s=2$, the system can be mapped onto the spectrum of a particle in a box. These are the only two interesting physical systems where the spectrum corresponds to the eigenvalues of a quadratic Hamiltonian. For our purposes it is sufficient to use the single-particle spectrum $\{\epsilon_m=m^s\}$ without further specifying the physical system, as was done in Ref.\cite{bib:muoi}.    

The total energy $E=n$ of the system may be distributed amongst ideal bosons in many different ways, so that 
\begin{equation}
    n=m_1^s+m_2^s+\cdots m_{\ell}^s\,=\,\sum_{i=1}^{\ell} m_i^s \,,
    \label{eq:n-summand-k}
\end{equation}
where ${\ell}$ is the maximum number of particles allowed, and the $m_i^s$ in the summand are all positive integers since the spectrum has only positive energy levels and the excitation energy is also positive. Such a summand always exists, but there are restrictions on the minimum number of terms for a given exponent $s$ depending on the value of $n$. 

The partition function of such a bosonic system is given by 
\begin{equation}
    Z(\beta)=\sum_n P^{(s)}(n) e^{-\beta\epsilon_n}\, =\,\prod_{m=1}^\infty \frac{1}{1-e^{-\beta m^s}}\, ,
\end{equation}
where $P^{(s)}(n)$ is the degeneracy of the state with energy $\epsilon_n$. As above, the energy is a dimensionless positive number, and $\beta$ is a positive dimensionless constant. This can always be achieved for a physical system since $\beta\epsilon_n$ is dimensionless. The $Z(\beta)$ is identical to the generating  function of number partitions with the identification $x=e^{-\beta}$. 

Following the method outlined in the appendix (\ref{appendix1}), the entropy is given by 
\begin{equation}
    S(\beta)=\beta E+\ln Z(\beta)\,=\, \beta E - \sum_{m=1}^{\infty}\ln \left(1-e^{-\beta m^s}\right)\,.
\end{equation}
Evaluating the sum over $m$ using the Euler-Maclaurin summation formula (see appendix \ref{appendix1} for details), we get 
\begin{equation}    
    S(\beta)=\beta E +\frac{1}{\beta^{1/s}}C_b(s) +\frac{1}{2}\ln\beta-\frac{s}{2}\ln 2\pi +O(\beta)\, ,
\end{equation}
where 
\begin{equation}
    C_b(s)= \Gamma(1+1/s)\zeta(1+1/s)\, .
\end{equation}
and $\zeta(x)=\sum_{j=1}^\infty \frac{1}{j^{\,x}}$  is the Riemann zeta function.
The saddle point is found by setting the first derivative $S_1(\beta)$ to zero, i.e.,
\begin{equation}
    S_1(\beta)=E-\frac{1}{s} \frac{C_b(s)}{\beta^{1+1/s}}+\cdots = 0\,,
\end{equation}
where we have neglected terms of $O(1/\beta)$ to get 
\begin{equation}
    \beta_0=\left(\frac{C_b(s)}{sE}\right)^{s/(s+1)}= \kappa_s E^{-\frac{s}{s+1}}\,.
\end{equation}
where $\kappa_s= (C_b(s)/s)^{s/(s+1)}$. Collecting the terms, we write the asymptotic density as
\begin{equation}
\overline\rho^{(s)}(E) = \frac{e^{S(\beta_0)}}{\sqrt{2\pi S_2(\beta_0)}}= \frac{\kappa_s}{(2\pi)^{(s+1)/s}}\sqrt{\frac{s}{s+1}} E^{-\frac{3s+1}{2(s+1)}}\exp\left[\kappa_s(s+1)E^{1/(s+1)}\right],
\label{eq:rholeading}
\end{equation}
where $E$ takes continuous values. For integer-valued $E$ we have
\begin{equation}
    P^{(s)}(n)=\rho^{(s)}(E=n)\,,
\end{equation}
where $P^{(s)}(n)$ is the number partition of $n$ into sums of integers raised to the power $s$. 

We first consider the results for $s=1$ in the unrestricted case. This is the most frequently analysed case, and the results are well known both asymptotically \cite{bib:ramanujan} as well as exactly in a series representation \cite{bib:rademacher}. The asymptotic result given in Eq.\ (\ref{eq:rholeading}) 
reduces to 
\begin{equation}
    \overline{\rho}^{(1)}(E)= \frac{1}{4E\sqrt{3}}\exp\left[\pi \sqrt{\frac{2E}{3}}\,\right]\,.
    \label{eq:rhos-1}
\end{equation}
and $P^{(1)}(n)=P(n)=\overline{\rho}^{(1)}(E=n) $. This is the famous Hardy-Ramanujan formula \cite{bib:ramanujan} published in the year 1917. It is interesting to note that the generating function used by Hardy and Ramanujan is in fact the partition function of a system of identical bosons. This was written many years before the seminal work of Bose published in 1924 \cite{bib:bose} which led to the birth of quantum statistics. 

As has been known from earlier work (see e.g. Ref. \cite{bib:muoi}), for partitions of positive integers into sums of integers the asymptotic expressions are approached very soon, i.e., for $n$ of the order of $100$ or more. This is so because in the case of integer partitions, as shown by Rademacher \cite{bib:rademacher}, the exact expression for integer partitions may be written as a convergent series. The $l$-th term in the series is of order $\exp(\frac{\pi}{l}\sqrt{\frac{2n}{3}})$. The leading term with $l=1$ gives the Hardy-Ramanujan result. The first correction to the exponent is at $l=2$
\begin{equation}
P(n)\approx \frac{1}{4E\sqrt{3}}\exp\left(\pi \sqrt{\frac{2n}{3}}\right)\left[1+c(n)
\exp\left({-\frac{\pi}{2}\sqrt{\frac{2n}{3}}}\right)\right]\,,
\label{eq:bosonscor-1}
\end{equation}
where $c(n)$ is an $n$-dependent prefactor. The correction to the leading asymptotic term falls off exponentially.

The $s=2$ unrestricted case  is also deduced easily from Eq.\ (\ref{eq:rholeading}). The asymptotic density of states is given by 
\begin{equation}
    \overline{\rho}^{(2)}(E) =\sqrt{\frac{2}{3}}\frac{\kappa_2}{(2\pi)^{3/2}E^{7/6}}
\exp\left[3\kappa_2E^{1/3}\right],
\end{equation}
where $\kappa_2=[\Gamma(3/2)\zeta(3/2)/2]^{2/3}\approx 1.10247...$. As before, 
$P^{(2)}(n)=\overline{\rho}^{(2)}(E=n)$, where $n$ is a positive integer. 

\begin{figure}[h]
    \centering
    \includegraphics[width=0.8\linewidth]{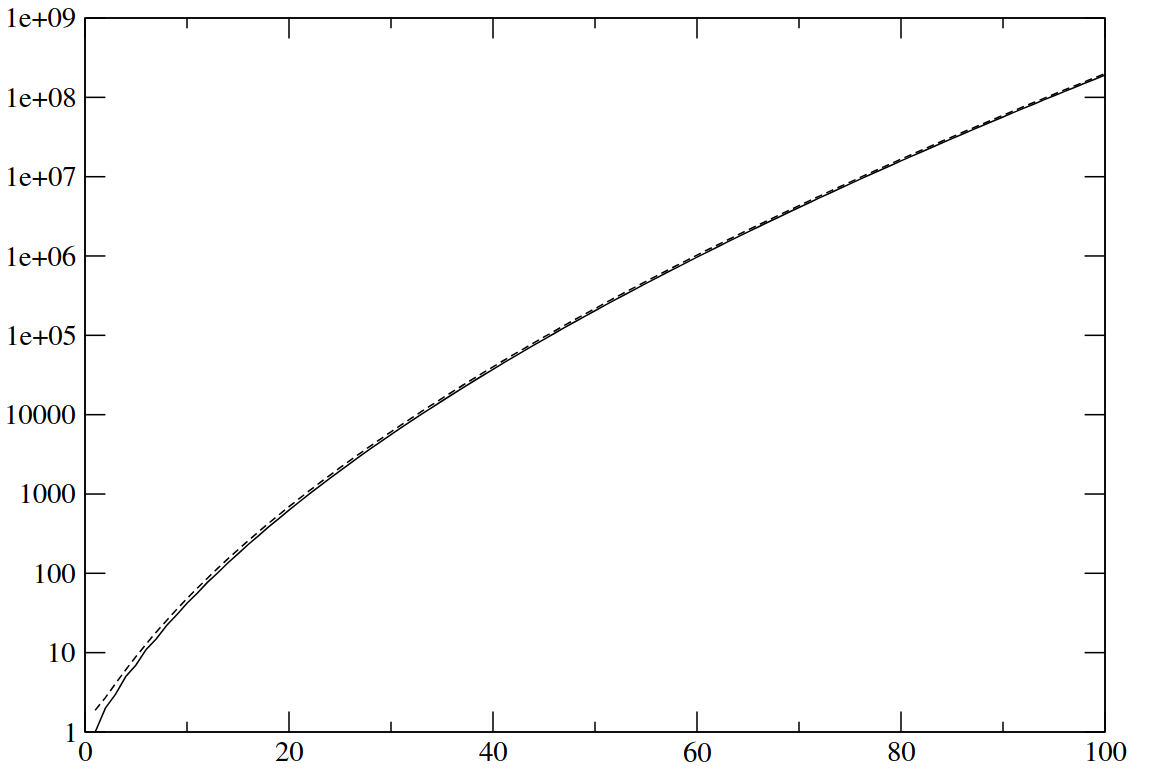}
    \put(-200,-10){n (or E)} 
    \put(-415,150){$P(n)$}
    \put(-415,130){$\overline{\rho}(E)$}
    \caption{Comparison of the exact P(n) (solid line) and the asymptotic density of states $\overline{\rho}(E)$, where E is a continuous parameter (dashed line). Here and in all other following figures, we have joined the points corresponding to $n$ by a continuous curve for comparison with the asymptotic behaviour. The error decreases exponentially as $E$ increases so that the asymptotic density approaches the exact density for all $E=n$.} 
    \label{fig:bosons-1}
\end{figure}

\begin{figure}[h]
    \centering
    \includegraphics[width=0.8\linewidth]{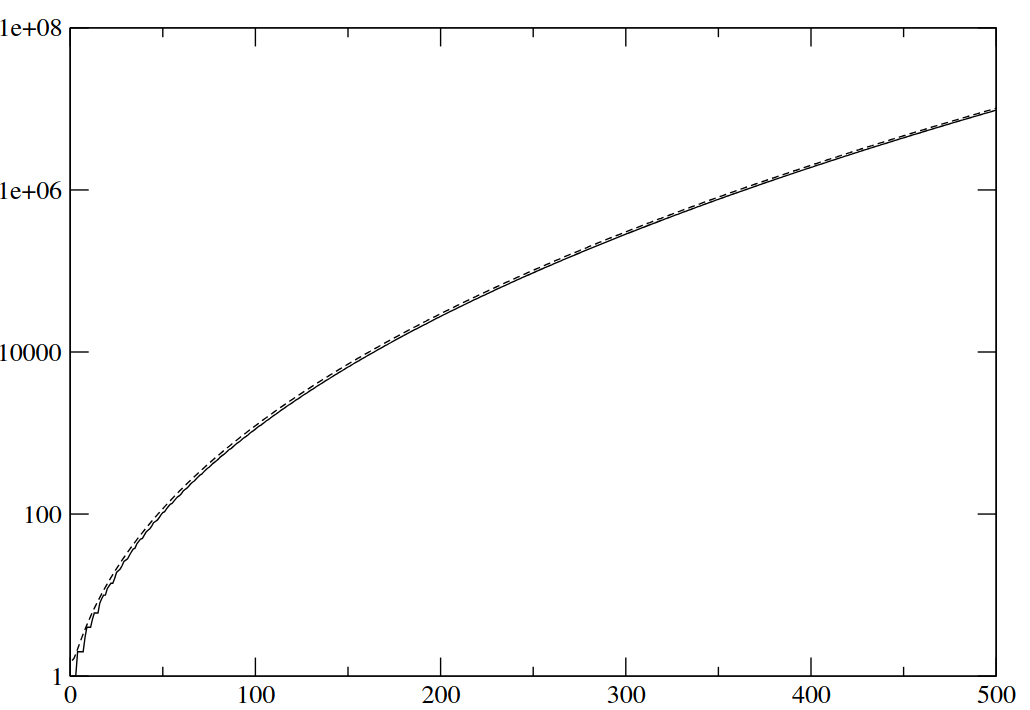}
    \put(-200,-10){n (or E)} 
    \put(-415,170){$P^{(2)}(n)$}
    \put(-415,150){$\overline{\rho}^{(2)}(E)$}
    \caption{Comparison of the $s=2$ exact $P^{(2)}(n)$ (solid line) and the asymptotic density of states $\overline{\rho}^{(2)}(E)$, where E is a continuous parameter (dashed line).} 
    \label{fig:bosons-2}
\end{figure}

Numerical results for the bosonic $s=1,2$ are shown in Fig.\ \ref{fig:bosons-1} and Fig.\ \ref{fig:bosons-2}, respectively. As expected, the asymptotic formula works well for $s=1$. In the case of $s=2$ it approaches more slowly, essentially controlled by the power of $E$ in the exponent. 

\subsection{Distinct partitions}

We next consider the case of distinct partitions. Unlike the bosonic or unrestricted partitions, where the connection is precise, the distinct partitions correspond to the fermionic case, but with a {\it{caveat}}. This is because the partition function of a fermionic system, as used by Bethe in deriving the nuclear level density, consists of $N$ spinless fermions distributed over the single-particle energy levels following the Pauli principle. The partition function of such a fermionic system, in the statistical Fermi gas model, is given by 
\begin{equation}
    Z(\beta) = e^{-\beta(N)(N+1)/2}\sum_n D(n) e^{-\beta\epsilon_n}\, =
               \,e^{-\beta(N)(N+1)/2}\prod_{m=1}^\infty \frac{1}{1-e^{-\beta m}}\, .
\end{equation}
Notice that, apart from the prefactor which depends on the ground state energy, the partition function is exactly the same as in the $s=1$ bosonic case. This leads to the same asymptotic behaviour as bosons with a different prefactor as seen in Eq.\ (\ref{eq:bethe-nucleus}). The main difference is due to the fact that the nucleons are fermions. The distribution of energy among fermions comes with the restriction imposed by the Pauli exclusion principle. The route to the asymptotic level density is the same as in the case of bosons with the above {\it proviso}. 

Distinct partitions with occupancy $k=1$ do not have any restriction on the ground-state distribution. In fact, they are identical to those of bosons, but the particle distribution follows the Pauli exclusion principle. The partition function of such a "fermionic" system in an arbitrary monomial potential is given by 
\begin{equation}
    Z(\beta)=\sum_n D^{(s)}(n) e^{-\beta\epsilon_n}\, =\,\prod_{m=1}^\infty{[1+e^{-\beta m^s}]}\, ,
    \label{eq:zfermi}
\end{equation}
where $D^{(s)}(n)$ is the degeneracy of the energy state $\epsilon_n$. The product converges absolutely if $0\le x < 1$. 

Using the general results outlined in the appendix \ref{appendix1}, the asymptotic many-particle density of states for a general spectrum specified by $s$ is given by
\begin{equation}
   \overline \rho_{f}^{(s)}= \sqrt{\frac{s\lambda_s}{4\pi(s+1)E^{(2s+1)/(s+1)}}}
\exp\left[\lambda_s(s+1)E^{1/(s+1)}\right],
    \label{eq:rhoofs}
\end{equation}
where $E$ takes continuous values. The subscript $f$ denotes the distinct ("fermionic") case as opposed to the unrestricted case. The constant $\lambda_s$ is given by 
$$\lambda_s= \left[\frac{1}{s}\Gamma(1+1/s)\eta(1+1/s)\right]^{s/(s+1)}\, , $$
where $\eta(s)=\sum_{l=1}^\infty(-1)^{(l-1)}\frac{1}{l^s}$ is the alternating zeta function. For integer $E$ we have, as before,
\begin{equation}
    Q^{(s)}(n)=\rho^{(s)}_f(E=n)\, ,
\end{equation}
where $Q^{(s)}(n)$ is the number of distinct partitions of $n$ into sums of integers raised to the power $s$. Though the asymptotic expression for $P^{(s)}(n)$ is known, the general expression for $Q^{(s)}(n)$ was derived only in \cite{bib:muoi}, surprisingly enough since the number partition is an old problem dating back to the time of Euler.  

\begin{figure}[h]
    \centering
    \includegraphics[width=0.7\linewidth]{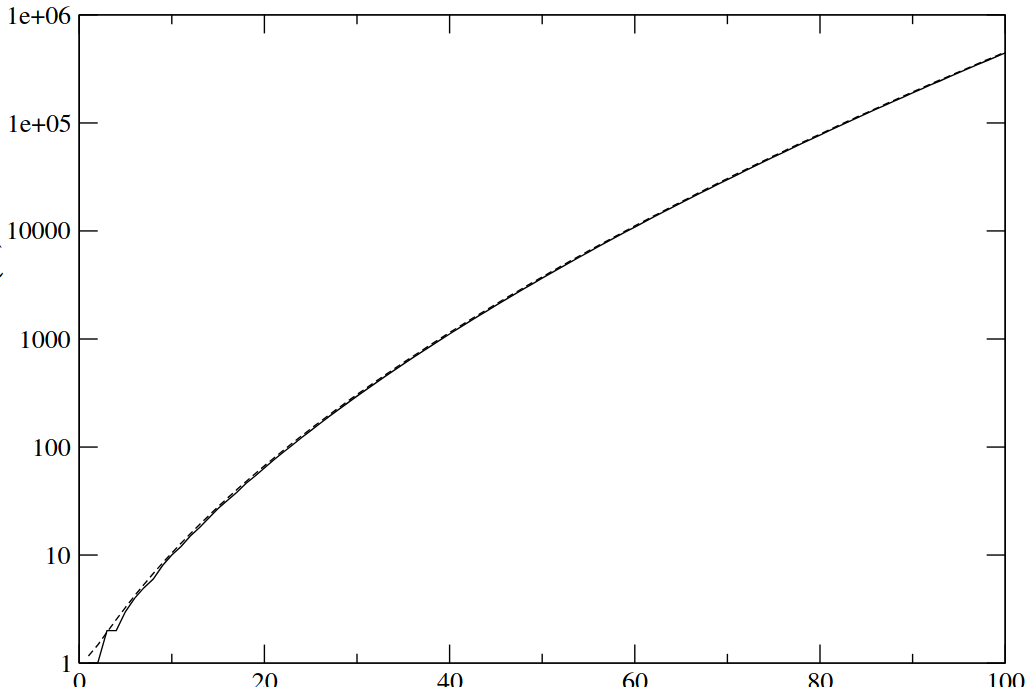}
    \put(-200,-10){n (or E)} 
    \put(-380,150){$Q(n)$}
    \put(-380,130){$\overline{\rho}_f(E)$}
    \caption{Comparison of the exact Q(n) (solid line) and the asymptotic density of states              $\overline{\rho}(E)$, where E is a continuous parameter (dashed line). The   error decreases exponentially as $E$ increases so that the asymptotic density              approaches the exact density for all $E=n$.} 
    \label{fig:fermions-1}
\end{figure}

We first consider the result for the $s=1$ case which again is well known \cite{bib:abramowitz}.  The asymptotic result given in Eq.\ (\ref{eq:rhoofs}) reduces to the well-known formula \cite{bib:abramowitz}
\begin{equation}
    \overline{\rho}_f(E)= \frac{1}{4\times3^{1/4}\times E^{3/4}}\exp\left[\pi \sqrt{\frac{E}{3}}\,\right]\,.
    \label{eq:rhos-f1}
\end{equation}
In Fig.\ \ref{fig:fermions-1}, we show explicitly a comparison of $Q^{(s)}(n)$ and the continuous $\overline{\rho}^{(s)}(E)$ for $s=1$. For the sake of comparison, the points corresponding to exact results at integer $n$ are joined by a continuous solid line.  

\begin{figure}[h]
    \centering
    \includegraphics[width=0.75\linewidth]{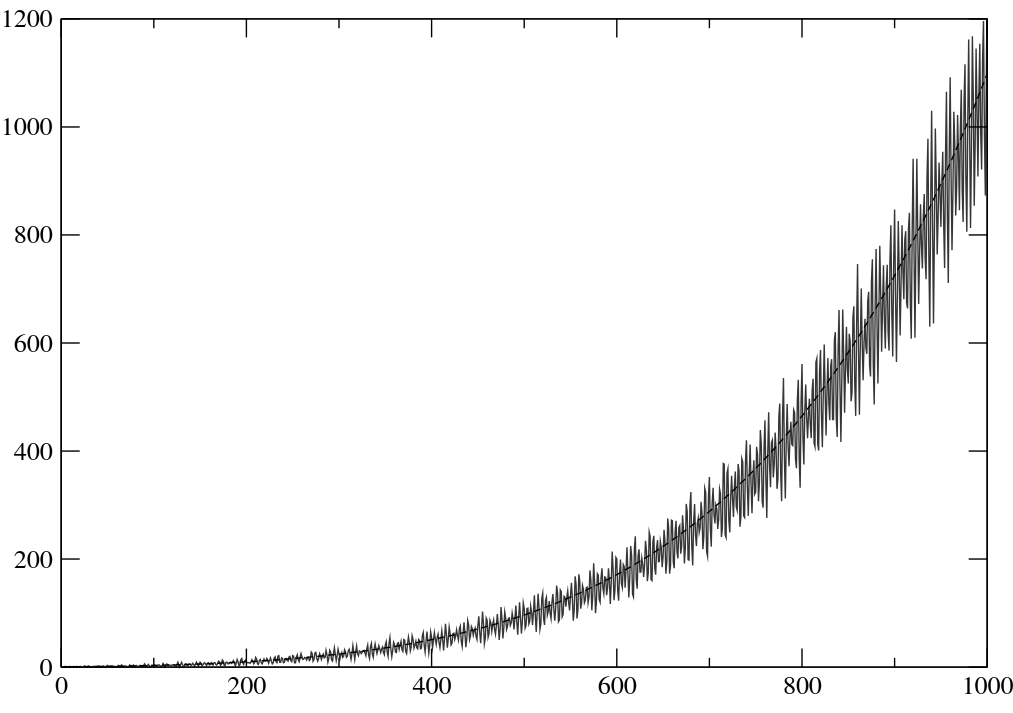}
    \put(-200,-10){n (or E)} 
    \put(-400,170){$Q^{(2)}(n)$}
    \put(-400,150){$\overline{\rho}_f^{(2)}(E)$}
    \caption{Comparison of the exact $Q^{(2)}(n)$ (solid line) and the asymptotic density of states 
$\overline{\rho}_f^{(2)}(E)$, where E is a continuous parameter (dashed line). } 
    \label{fig:fermions-2}
\end{figure}

For distinct square partitions with $s=2,\, k=1$, the leading asymptotic density is given by
\begin{equation}
    \overline{\rho}_f^{(2)}(E)= \sqrt{\frac{\lambda_2}{6\pi}}\,E^{-5/6}\,
                  \exp\left(3\lambda_2 E^{1/3}\right),
    \label{eq:rhosf-2}
\end{equation}
where $\lambda_2=0.486227919$. A comparison of the exact and asymptotic results up to  $n=1000$ is shown in Fig.\ \ref{fig:fermions-2}. 
The figure reveals that the asymptotic function,  represented by the solid line, does not follow the average value of the exact $Q^{(2)}(n)$, unlike for the other partitions discussed above. This means that the asymptotic is yet to be reached. 

Considerably improved average values are obtained by adding corrections to the leading saddle-point approximation of the smooth part of the level density. Following the method discussed in section {\ref{sec:spcor}} and Eq.\ (\ref{eq:asymsp11}), we can derive higher-order saddle-point contributions, yielding a better asymptotic form $\overline\rho_f^{(2)}(E)$ up to third-order, giving
\begin{equation}
\overline{\rho}_{fc}^{(2)}(E)= \sqrt{\frac{\lambda_2}{6\pi}}\,E^{-5/6}\,e^{3\lambda_2E^{1/3}}
            \left[1-c_1\, E^{-1/3}-c_2\, E^{-2/3}-c_3\, E^{-1}\right],
\label{eq:rhof2cor}
\end{equation}
with the coefficients $c_1=0.285645648$, $c_2=0.057115405$ and $c_3=0.020665371$ as explicitly calculated in Ref.\cite{bib:mbbb}.

\begin{figure}[h]
    \centering
    \includegraphics[width=0.75\linewidth]{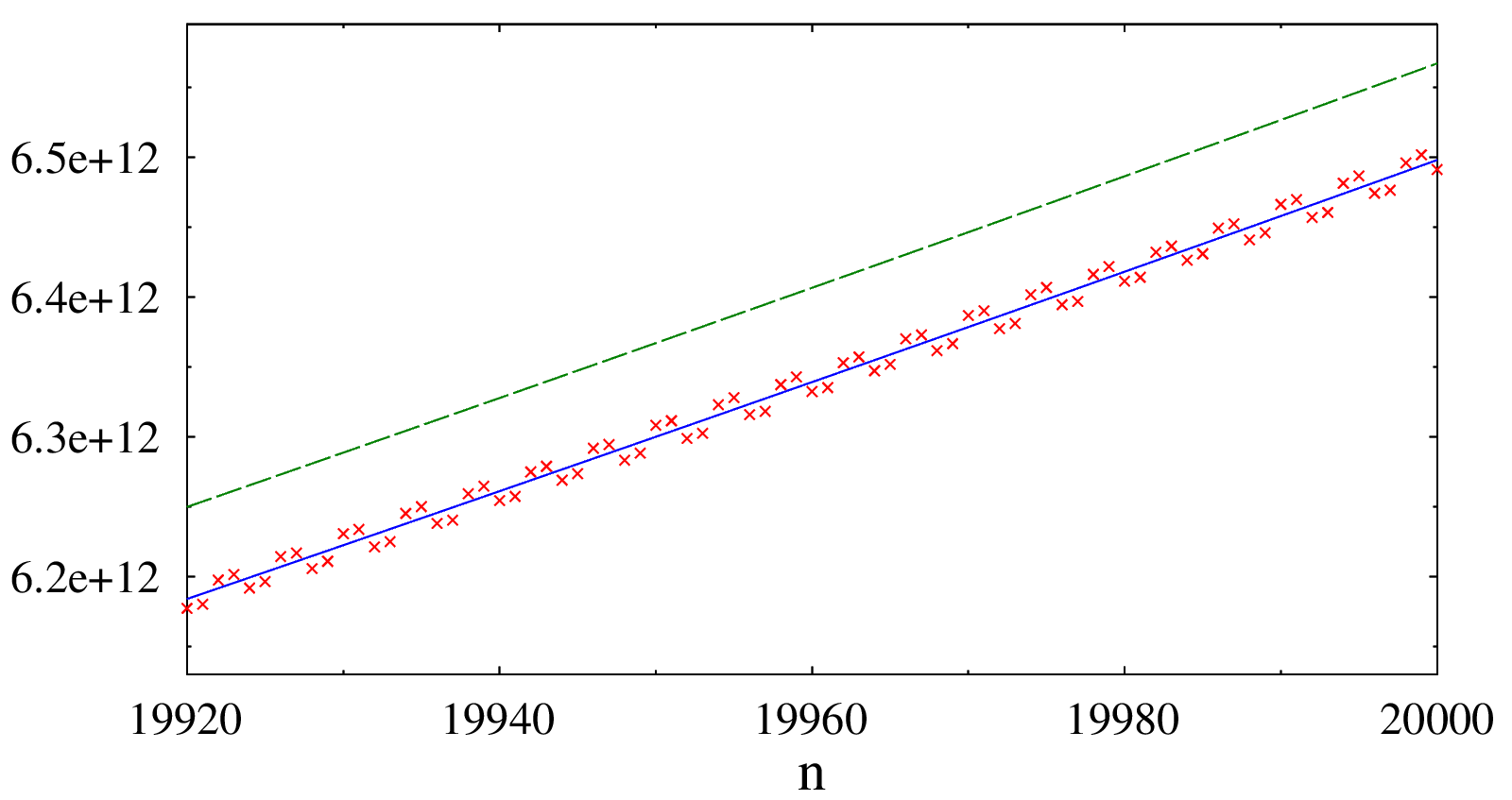}
    \put(-160,3){= E} 
    \put(-400,140){$Q^{(2)}(n)$}
    \put(-400,120){$\overline{\rho}_{fc}^{(2)}(E)$}
    \caption{Comparison of the exact $Q^{(2)}(n)$ (points-red) and the asymptotic density of states 
             $\overline{\rho}_{fc}^{(2)}(E)$, where E is a continuous parameter (green-dashed line). 
             The corrected asymptotic density is also shown (blue-continuous line). } 
    \label{fig:fermions-2sp}
\end{figure}

The improvement with the corrections to the leading asymptotic part can be seen in Fig.\ \ref{fig:fermions-2sp}, where we recognize that the corrected asymptotic density in Eq.\ (\ref{eq:rhof2cor}) represents the average much better than the leading asymptote (\ref{eq:rhosf-2}). 

It was pointed out in \cite{bib:mbbb} that the exact $Q^{(2)}(n)$ of the distinct square partitions exhibits pronounced regular oscillations with a beat structure. In any partition of $n$, for smaller values, we expect irregular variations since we are dealing with integers expressed in terms of other integers. As $n$ becomes large, these variations go towards zero exhibiting a smooth asymptotic limit. However, in the case of the distinct square partitions, $Q^{(2)}(n)$, regular oscillations persist even for very large $n$. Indeed, there is a beat-like structure oscillating around the average asymptotic function $\overline\rho_f^{(2)}(E)$, as seen in Fig.\ \ref{fig:fermions-2} where the exact partitions are joined by a continuous smooth curve.  

Where do these {\it regular oscillations}, including the beat-like structure, come from? There is no physical scale (needed for beats) in the problem that can give raise to frequencies. To look at its origins, we give here a qualitative argument to set the stage for a more quantitative analysis in the next section. Going back to Eq.\ (\ref{eq:zfermi}), we have
\begin{equation}
    \ln Z(\beta)\,=\,\sum_{m=1}^\infty \ln\left[1+e^{-\beta m^2}\right]\,=\,
                   \int_1^\infty d\epsilon\,g(\epsilon)\ln\left[1+e^{-\beta\epsilon}\right],
    \label{eq:zbetaf-2}
\end{equation}
where the single particle density $g(\epsilon)$ is given by
\begin{equation}
    g(\epsilon)\,=\, \sum_{m=1}^{\infty} \delta(\epsilon-m^2) \,=
                  \,\frac{1}{2\sqrt{\epsilon}}\left[1+2\sum_{k=1}^\infty 
                  \cos(2\pi k\sqrt{\epsilon})\right].
    \label{eq:gofedistinct}
\end{equation}
Substituting Eq.\ (\ref{eq:gofedistinct}) into Eq.\ (\ref{eq:zbetaf-2}), the integral may be evaluated exactly, giving
\begin{equation}
  \ln Z(\beta)\,=\,\frac{C}{\sqrt{\beta}}-\frac{1}{2}\ln(2)+\sqrt{\frac{\pi}{\beta}}\sum_{\ell=1}^\infty 
                 \frac{(-1)^{\ell+1}}{\ell^{3/2}}\sum_{q=1}^\infty e^{-\pi^2q^2/\ell\beta}\,;
                  ~~~C=\Gamma(3/2)\eta(3/2)\,,
  \label{eq:Zbetaf-2x}
\end{equation}
where $\eta(x)$ is the Dirichlet eta function and the entropy is given as usual by $S=\beta E+\ln Z$.  The many-particle density is obtained in principle by Laplace inverting the partition function as outlined in Sec.\ \ref{sec:mpds}. In practice, however, it is almost impossible to invert this highly non-linear equation. Instead, we use the saddle-point approximation to obtain the smooth asymptotic behaviour which results from the first two terms in Eq.\ (\ref{eq:Zbetaf-2x}). Under this saddle-point approximation, we may evaluate the contribution of third term in Eq.(\ref{eq:Zbetaf-2x}) to the density. We find 
\begin{equation}
    \delta \rho^{(2)}_f(E)\,\propto\, \frac{1}{\sqrt{E}}\sum_{\ell=1}^\infty 
    \frac{(-1)^{\ell+1}}{l^{3/2}}\sum_{q=1}^\infty \,\cos[2\pi q\sqrt{(E/\ell)}]\,, 
\end{equation}
which in principle can lead to regular oscillations, and even to beats. In particular, when the energy value is such that $\sqrt{E/\ell}=m_\ell$, where $m_\ell$ is a positive integer, we get
\begin{equation}
    \delta \rho^{(2)}_f(E)\,\propto\, \sum_{\ell=1}^\infty \frac{(-1)^{\ell+1}}{l^2m_\ell}\sum_{q=1}^\infty 
    \,\cos[2\pi qm_\ell]\,. 
\end{equation}
This particular case is interesting since the argument of the cosine, $2qm_\ell$ , forms a  Pythagorean triplet along with $(q^2+m_\ell^2), \, (q^2-m_\ell)^2$.

Two crucial points emerge from this very qualitative analysis. First, the regular oscillations seen in the exact partitions are a direct result of the oscillatory behaviour of the density. Second, the occurrence of Pythagorean triplets may have some thing to do with regular oscillations and beats.  We provide a quantitative analysis in the next section, where we develop a semiclassical trace formula for the oscillations in the distinct square partitions. 

\section{Quantitative semiclassical analysis of distinct square partitions}
\label{sec:semiclass}

Until now we have only been comparing the exact partitions with their asymptotic smooth behaviour. The results are well known for specific cases, but the general result for arbitrary spectrum with power $s$ and level occupation $k$ obtained in Eq.\ (\ref{eq:rhofty1}) was derived for the first time in Ref.\cite{bib:kutta}.  In the present section we shall focus our attention on the distinct square partitions where, as stated earlier, there occur regular oscillations with a beat structure that has not been observed in any other partitions.\footnote{We have already seen their onset for values 
$n\leq 1000$ in Fig.\ \ref{fig:fermions-2sp} above.} We shall present here only the most important ideas, definitions and results; full details are given in Ref.\ \cite{bib:mbbb}.

\subsection{Fourier analysis of the level density}
In order to obtain an understanding of the origin of the oscillations in $Q^{(2)}(n)$, we first study the Fourier Transform (FT) of the level density $\rho^{(2)}_F(E)$, which is given by
\begin{equation}
    F(\tau)=\int_{-\infty}^{+\infty} dE\, \rho^{(2)}_f(E)e^{-iE\tau}\,,
    \label{eq:ftrho}
\end{equation}
where $E$ and $\tau$ are a pair of conjugate dimensionless variables, say energy and time. The absolute value of the FT is 
\begin{equation}
    |F(\tau)|=\exp{Re[\ln\,Z(i\tau)]}\,,
    \label{eq:ftrhoabs}
\end{equation}
where $Z(\beta)$ is given by 
\begin{equation}
    Z(i\tau)=\exp\left\{\sum_{m=1}^M\,\ln[1+\exp(-m^2i\tau)]\right\}.
    \label{eq:distZ}
\end{equation}
In principle, the largest integer $M$ occurring in the partition of $n$ as a distinct sum of squares  is $\infty$. In practice, however, $M$ is the largest integer less than or equal to $\sqrt{n}$, i.e., $M=[\!\sqrt{n}\,]$. We now study the amplitude $|F(\tau)|$ as a function of the frequency $f=2\pi/\tau$.  For convenience in displaying the numerical results, we normalise the FTs by the maximum amplitude $I_0=|F(\tau)(2\pi k)|=\exp(M\ln2)$ ($k=0,1,2,\ldots,$). 

\begin{figure}[h]
    \centering
    \includegraphics[width=0.7\linewidth]{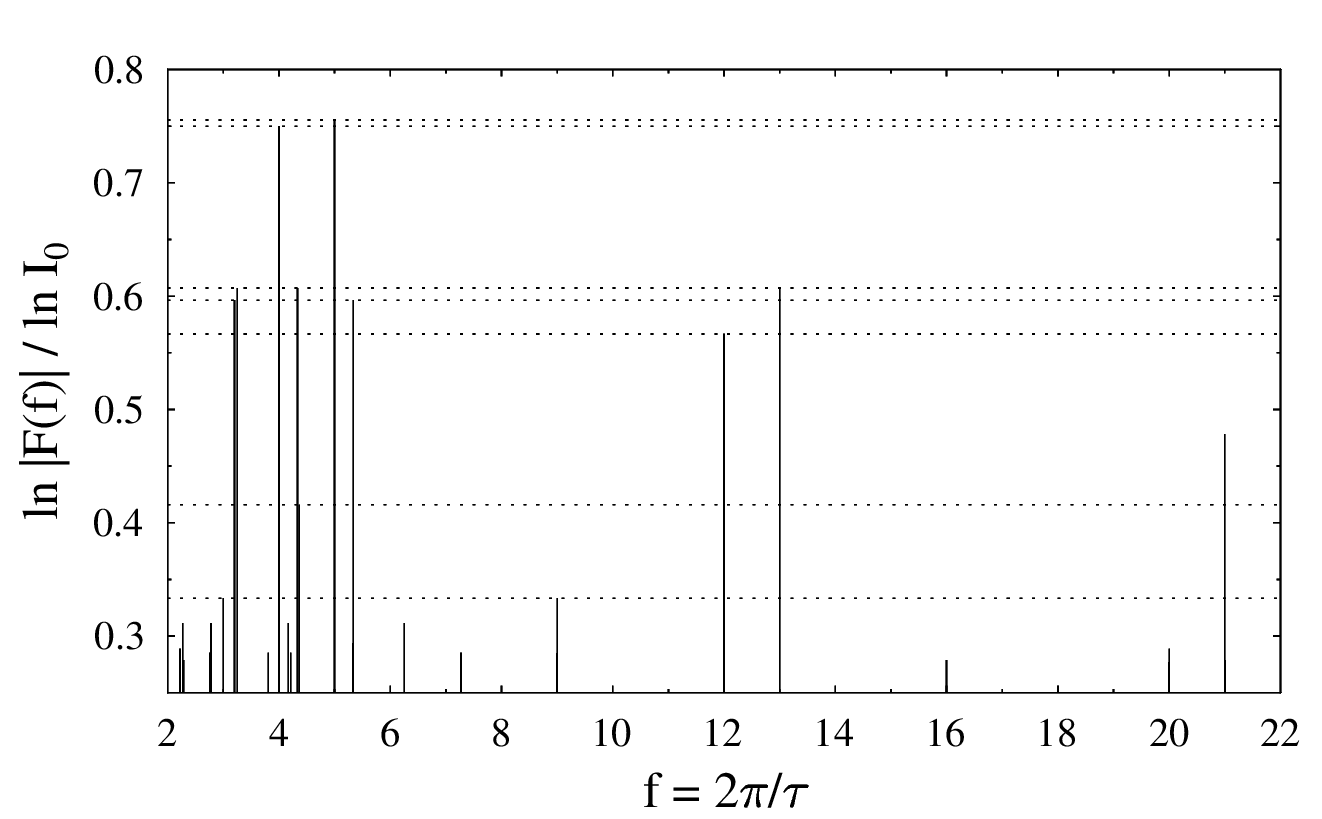}
    \includegraphics[width=0.7\linewidth]{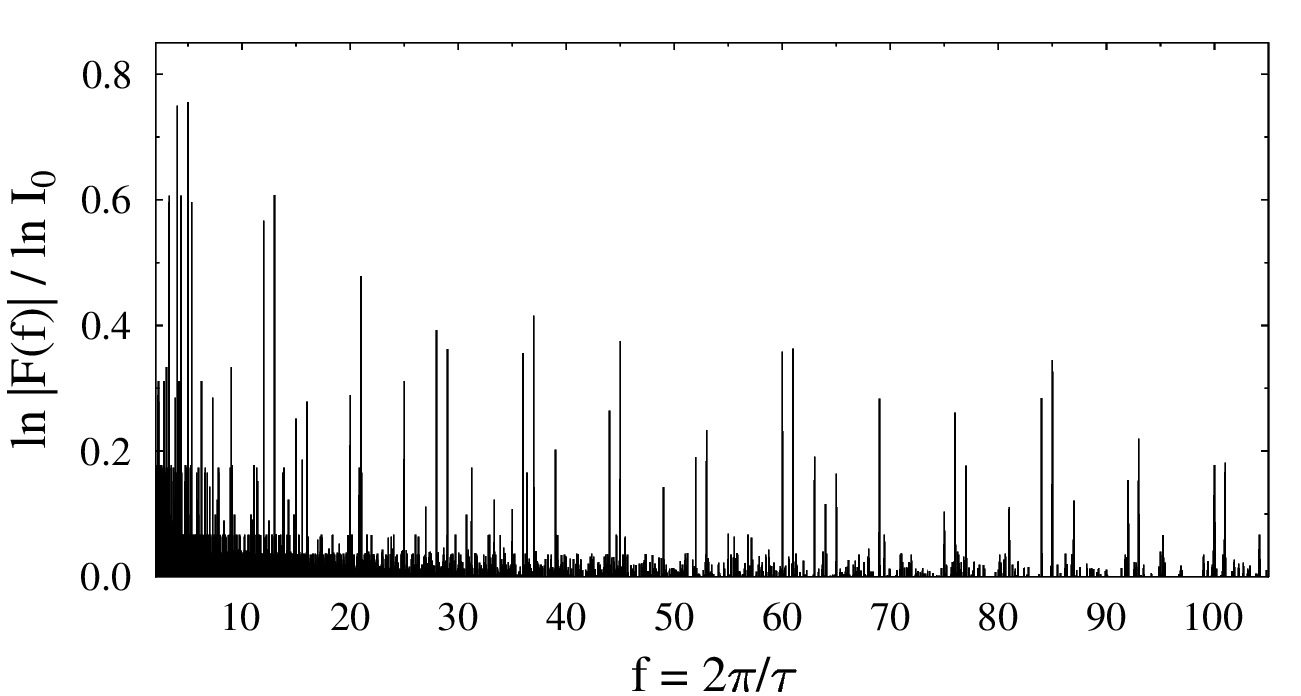}
    \caption{Scaled Fourier transform $\ln|F(f)|$ of level density on a log vertical scale.              The upper figure shows the peaks for $0\le f\le 22$. The horizontal dotted lines   separate the calculated relative intensities of the first 10 generations.             The lower panel shows the  Fourier peaks over a larger range of frequencies.} 
    \label{fig:Ft(22)}
\end{figure}

Figure \ref{fig:Ft(22)} shows the Fourier spectrum as a function of the frequency $f$. The top panel shows the Fourier peaks for $2\le f \le 22$ plotted on a log scale and the bottom panel shows the same for $f$ up to 105.   
We notice the sharp peaks located at the integer values of $f=3,4,5,9,12,13,16,20, 21$ in the upper panel. The horizontal lines, corresponding to scaled intensities (calculated as discussed in the next subsection), classify the peaks in terms of their intensities into {\it generations} (labeled $g$) with decreasing intensities. What is striking about the spectrum is the appearance of the two most prominent peaks at $f=4$ and 5, numbers which (together with $f=3$) appear in the smallest primitive Pythagorean Triplet (PT) $(3,4,5)$. Furthermore, the peaks at $f=12$ and 13 also appear as part of the second smallest primitive PT (5,12,13).  

To investigate this pattern further, we carefully look at the lower panel. We notice that there are many consecutive primitive pairs such as (4,5),\ (12,13),\ (28,29),\ (36,37),\ (45,46),\ (60,61),\ (84,85),\  and (100,101) in the range shown in the lower panel of Fig.\ \ref{fig:Ft(22)}.   Again, some of these pairs appear together in primitive PTs such as (11,60,61) and (13,84,85). Even when they are isolated like (28,29),\ (45,46) and 101, they appear in other PTs such as (28,45,53) or (20,21,29). Some other peaks appear in scaled PTs like (12,16,20). The hierarchies of these peaks are summarised in table (\ref{tab:hierarchy}) for the first 10 generations.
\begin{table}[]
    \centering
   \begin{tabular}{|c|c|c|c|c|c|c|c|}\hline
\;g\; &$\tau_g$ & $f_g$ & \,$\ln I_g/\!\ln I_0\,$ & $\Lambda_g$& $\mu_g$&$\kappa_g$ & $\phi_g$ \\ 
\hline
0 &0          & 1    & 1.0    &0.48622&0&3.085&0\\
1 &$2\pi/5$   & 5    & 0.7554 &0.37444&0&4.006&0\\
2 &$2\pi/4$   & 4    & 3/4    &0.40000&0.199919&3.352&-0.2318\\
3 &$2\pi/13$  & 13   & 0.6072 &0.30743&0&4.877&0\\
4 &$6\pi/13$  & 13/3 & 0.6072 &0.30743&0&4.877&0\\
5 &$8\pi/13$  & 13/4 & 0.6072 &0.30743&0&4.877&0\\
6 &$6\pi/16$  & 16/3 & 0.5964 &0.3200&-0.154191&4.219&0.2242\\
7 &$10\pi/16$ & 16/5 & 0.5964 &0.3200& 0.154191&4.219&-0.2242\\
8 &$2\pi/12$  & 12   & 0.5667 &0.3129&0.160611&4.264&-0.2358\\
9 &$22\pi/48$ & 48/11& 0.4158 &0,2450&-0.141520&6.502&0.2542\\
10&$2\pi/3$   & 3    & 0.3333 &0.2847&0.364242&3.244&-0.4538\\
\hline
\end{tabular} 
    \caption{Hierarchy of Fourier peaks, giving the periods, frequencies and scaled intensities      gleaned from Fig.\ \ref{fig:Ft(22)}, upper panel. The constants $\Lambda_g$, $mu_g$, $\kappa_g$      and $\phi_g$ corresponding to each generation $g$ are defined in connection with the trace      formula discussed in the next subsection.}
    \label{tab:hierarchy}
\end{table}

Thus, there appears a strong evidence that the PTs play a dominant role in the Fourier spectrum and hence in the oscillations of the level density. Indeed, this will be confirmed quantitatively by the semiclassical trace formula presented next. 

\subsection{A trace formula for the distinct square partitions}

In order to explain the oscillations in the distinct square partitions quantitatively, we sketch here the derivation of a semiclassical trace formula for the exact level density, following methods discussed in the textbook \cite{bib:book}, and present some of its results.
 
\begin{figure}[h]
    \centering
    \includegraphics[width=0.4\linewidth]{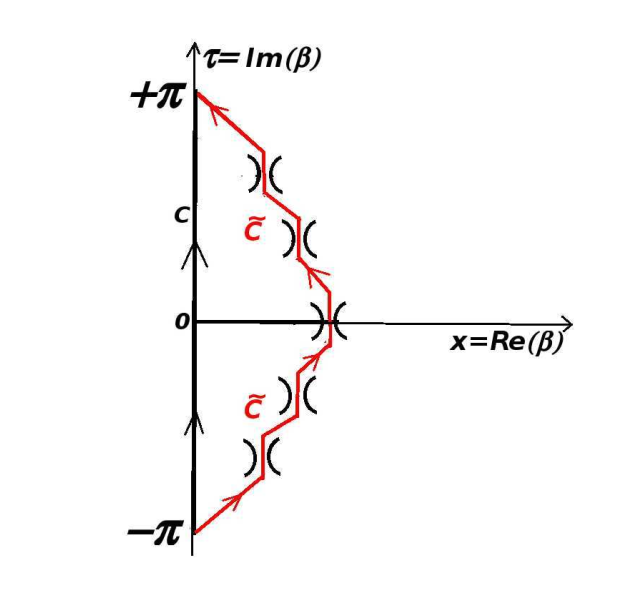}
    \caption{Schematic plot of the contour integral in Eq.\ (\ref{eq:d2ofn}). The exact contour $C$ is              deformed into a new contour $\tilde C$ passing over the most prominent saddles in the complex plane.}
    \label{fig:contour}
\end{figure}

It is known that asymptotic expressions of oscillating functions can be obtained using stationary-phase integration over saddles in the complex plane. A scholarly example of this procedure is given for the Airy function in Ref.\ \cite{bib:airy} (whose summary is found in Appendix B of Ref.\ \cite{bib:mbbb}). The idea of this approach is the following. We start from the exact integral representation of the partition $Q^{(2)}(n)$ as obtained from Eq.\ (\ref{eq:pofn}):
\begin{equation}
Q^{(2)}(n) = \frac{1}{2\pi} \int_{-\pi}^{\pi} {\cal R}e\,
         \exp\!\left[in\tau+\sum_{m=1}^{[\sqrt{n}\,]}
         \ln\left(1+e^{-im^2\tau}\right)\right]\!{\rm d}\tau
         \label{eq:d2ofn}
\end{equation}
This is an integral in the complex $\beta=\epsilon+i\tau$ plane yielding the exact $Q^{(2)}(n)$, whereby the integration contour $C$ is the line along the imaginary $\tau$ axis from $-\pi$ to $+\pi$. The exact partition function is obtained from integrating along this line, whereby the oscillations originate from the peaks along the time (i.e., period) axis $\tau$. Since the integrand of Eq.\ (\ref{eq:d2ofn}) has no singularities for Re$(\beta)>0$, we are allowed to deform the contour arbitrarily into the complex plane, keeping its end points fixed. Specifically, we choose a new contour $\widetilde C$ which passes over the most dominant saddles in the complex $\beta$ plane, as sketched in Fig.\ \ref{fig:contour}. Each of the chosen saddles emerges from a peak on the $\tau$ axis at the position $\tau_g$ of the Fourier peaks listed in Tab.\ \ref{tab:hierarchy}. We then do the stationary-phase integration locally at each saddle to obtain a contribution to the asymptotic (or semiclassical) expression for the oscillating partition. The exact path between the selected saddles does not matter, since only local contributions near the saddles are collected. 

The real saddle point at $\tau=0$ accounts for the smooth asymptotic partition in Eq.\ (\ref{eq:rhosf-2}), while all complex saddles account for its oscillating part which we define by
\begin{equation}
    \delta Q(n)=\rho_f^{(2)}(n)-\overline\rho_{fc}^{(2)}(n)=Q^{(2)}(n)-\overline Q^{(2)}(n); ~~E=n \, ,
    \label{eq:deltaD}
\end{equation}
where we have subtracted the corrected average part of the level density given in Eq.\ (\ref{eq:rhof2cor}) and $\overline Q^{(2)}(n)=\overline{\rho}_{fc}^{(2)}(E$=$n)$ denotes the distinct square partition in the asymptotic limit. The complex saddles with $\tau_g\neq 0$ thus contribute to $\delta Q(n)$ valid in the asymptotic limit of large $n$. These complex saddles must be found numerically for each generation $g$ and for each $n$ by scanning the landscape of Eq.\ (\ref{eq:rhosf-2}) in the complex plane. Analytical fits to the local properties of the saddles, with which the stationary-phase integrals are evaluated, have been given in Ref.\ \cite{bib:mbbb}. 

We reproduce now the result for $\delta Q(n)$ derived in \cite{bib:mbbb}. Summing over all saddles with $\tau_g\in(-\pi,+\pi$), $\tau_g\neq 0$, yields the {\it semi-classical trace formula}
\begin{equation}{\delta Q(n) = \sum_g A_g(n)
              \cos\!\left[n\,\tau_g-3\mu_g\,n^{1/3}+\varphi_g\right]}\, ,
              \label{eq:dgsc}
              \end{equation}
where the amplitudes are given by
$$A_g(n) = \frac{2}{(4\pi \kappa_g)^{1/2}}\,n^{-5/6}\,
         e^{3\Lambda_g\,n^{1/3}} \, .$$
Note that the amplitude $A_0(n)$ (corresponding to the real saddle at $\tau_0=0$ is identical to the asymptotic smooth part $\overline\rho^{(2)}_{f}(n)$ in Eq.\ (\ref{eq:rhosf-2}). The constants $\mu_g, \phi_g, \kappa_g$ and $\Lambda_g$ were determined numerically (see Ref.\ \cite{bib:mbbb} for details); their values are given in Tab.\ \ref{tab:hierarchy}. 
\begin{figure}[h]
\centering
\vspace*{-0.3cm}
\includegraphics[width=12cm,angle=0]{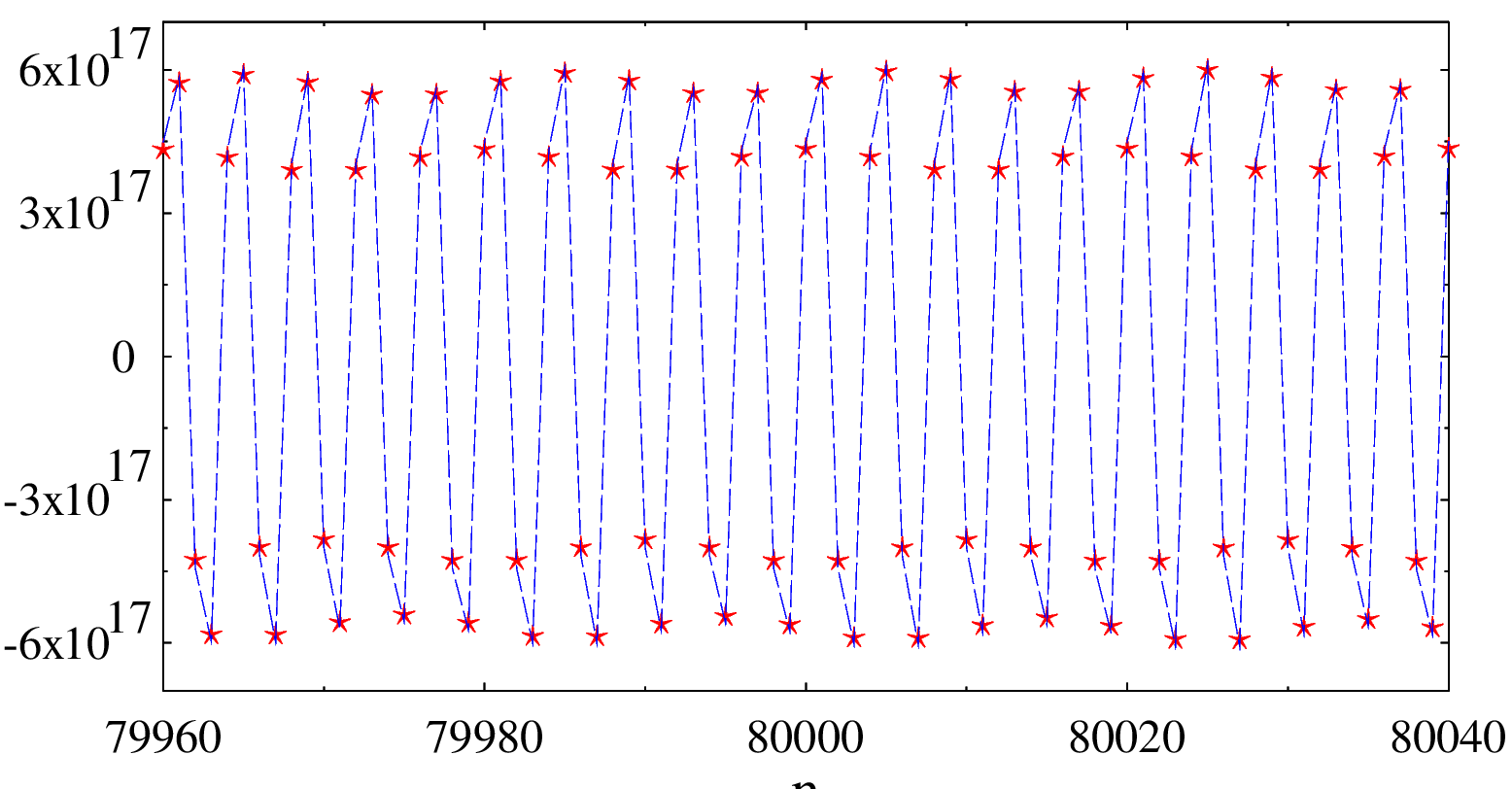}
\put(-370,100){$\delta Q(n)$}
\vspace*{-0.2cm}
\includegraphics[width=12cm,angle=0]{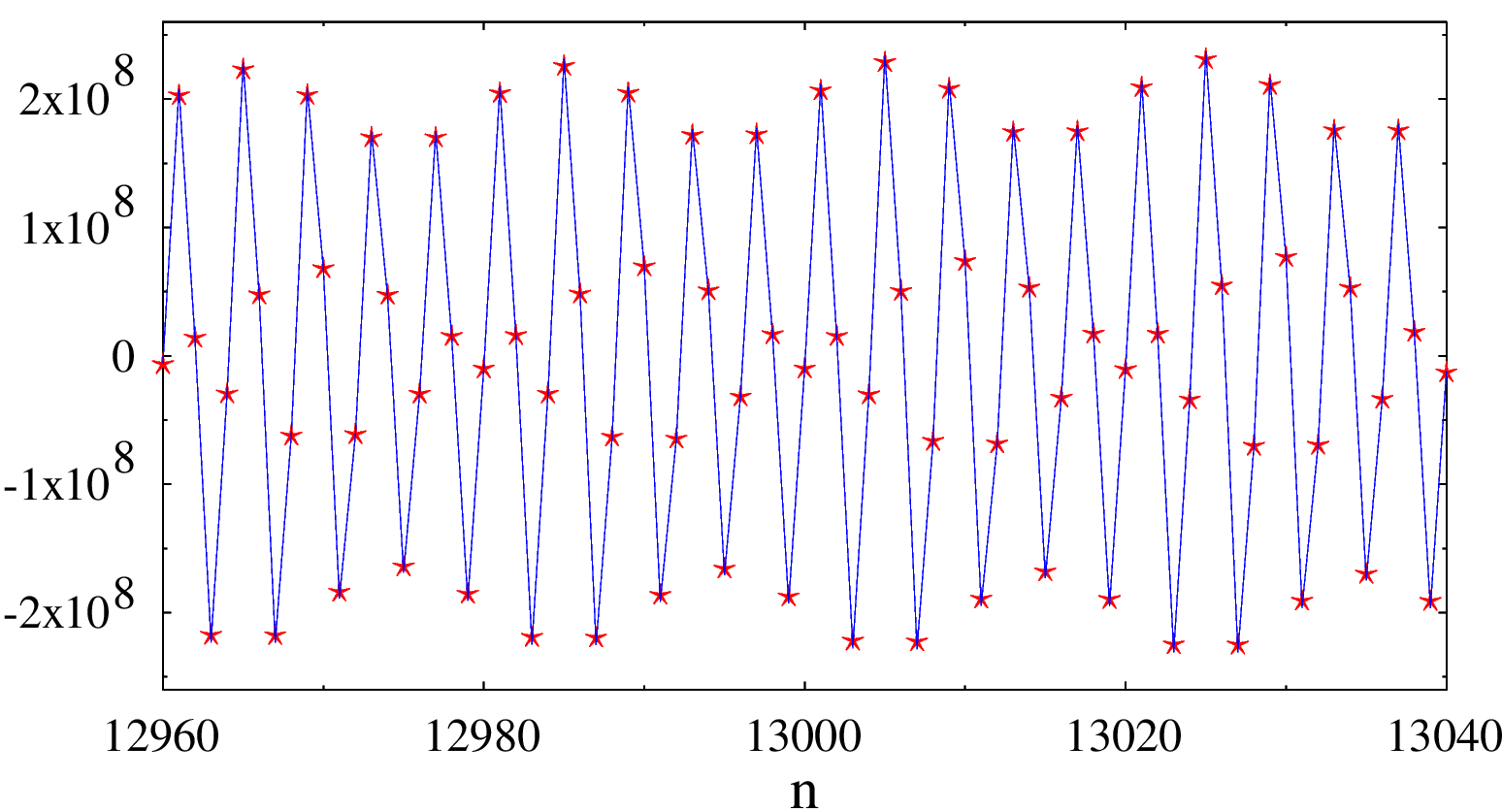}
\put(-370,100){$\delta Q(n)$}
\caption{Result of the trace formula Eq.\ (\ref{eq:dgsc}), shown by the dashed (blue) line, versus the exact $\delta Q(n)$, shown by the red stars, in two regions of large $n$.}
\label{fig:dp3}
\end{figure} 

\begin{figure}[h]
\centering
\vspace*{-0.3cm}
\includegraphics[width=12cm,angle=0]{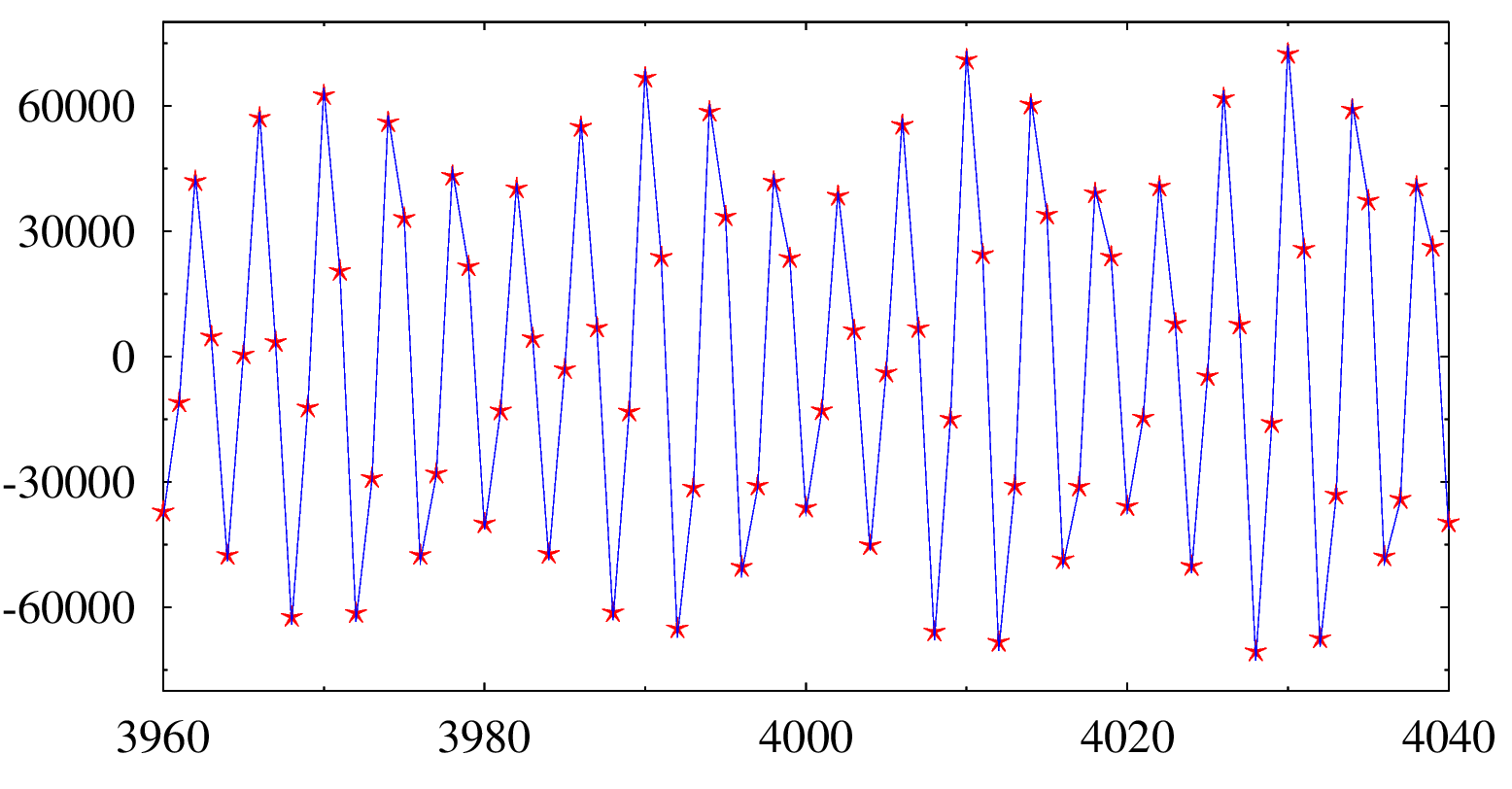}
\put(-370,100){$\delta Q(n)$}

\includegraphics[width=12cm,angle=0]{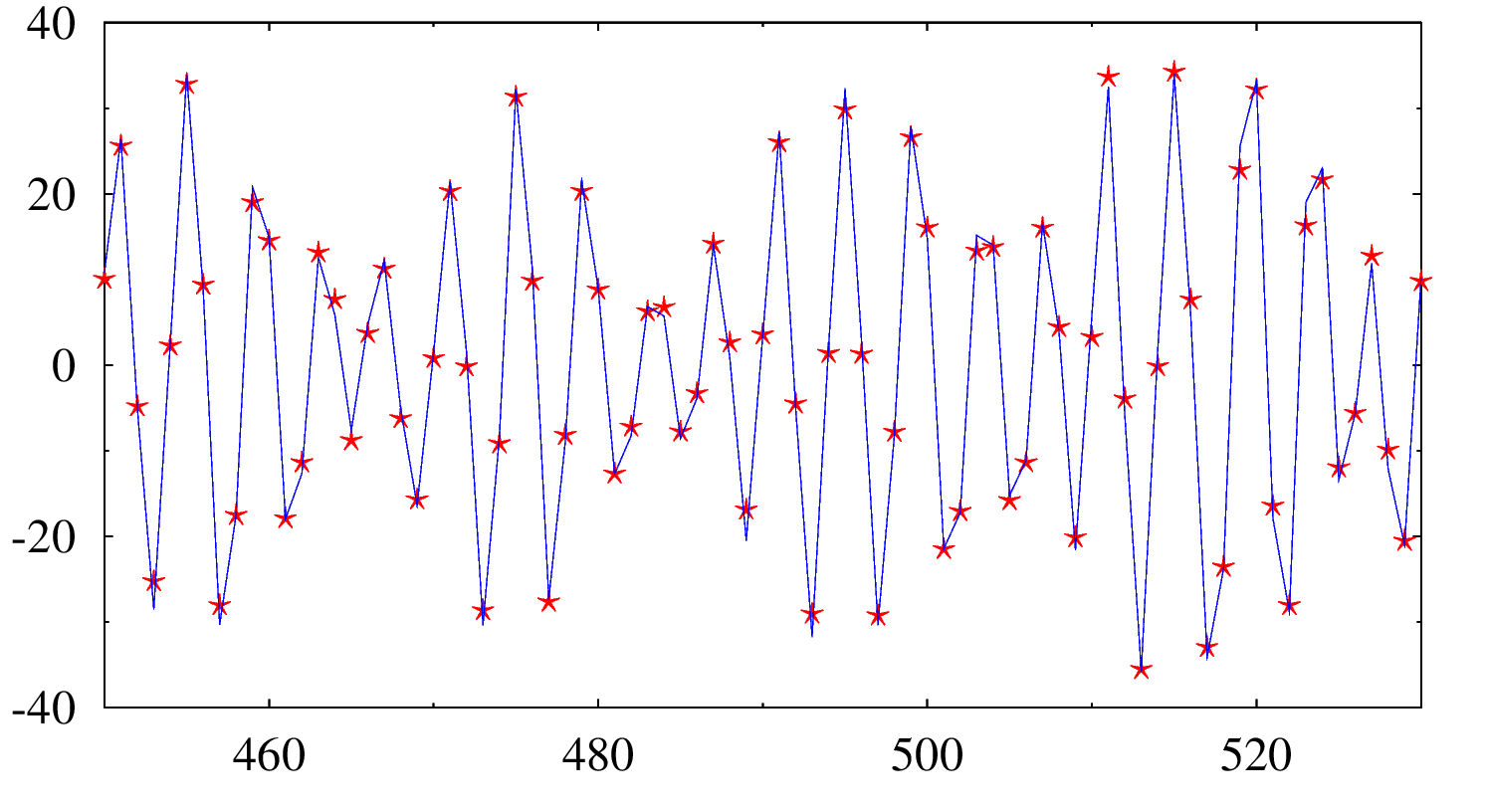}
\put(-370,100){$\delta Q(n)$}
\put(-170,5){$n$}
\caption{Result of the trace formula Eq.\ (\ref{eq:dgsc}), shown by the dashed (blue) line, versus          the exact $\delta Q(n)$, shown by the solid (red) line, in two regions of small $n$.}
\label{fig:dp4}
\end{figure}

Figs.\ \ref{fig:dp3} and \ref{fig:dp4} show the results of the trace  formula (\ref{eq:dgsc}) by blue lines, compared to the exact $\delta Q(n)=Q^{(2)}(n)-\overline Q^{(2)}(n)$ (solid lines, red) in four ranges of $n$. The agreement between the two curves is excellent in all regions of $n$, the semi-classical results reproducing perfectly both the rapid  oscillations and the beating amplitude of the exact $\delta Q(n)$.
In the calculations for these results, the generations $g=1$ - 9 have been included. However, nothing changes visibly in the results for $n\gtrapprox 4000$ if we only include the two leading generations 1 and 2. While this might be a surprise at first sight, it can be explained by the values of the constant $\Lambda_g$, which regulate the exponential growth of the amplitudes $A_g(n)$. These are clearly higher for generations 1 and 2 than for the others.

\begin{figure}[h]
\centering
\vspace*{-0.3cm}
\includegraphics[width=12cm,angle=0]{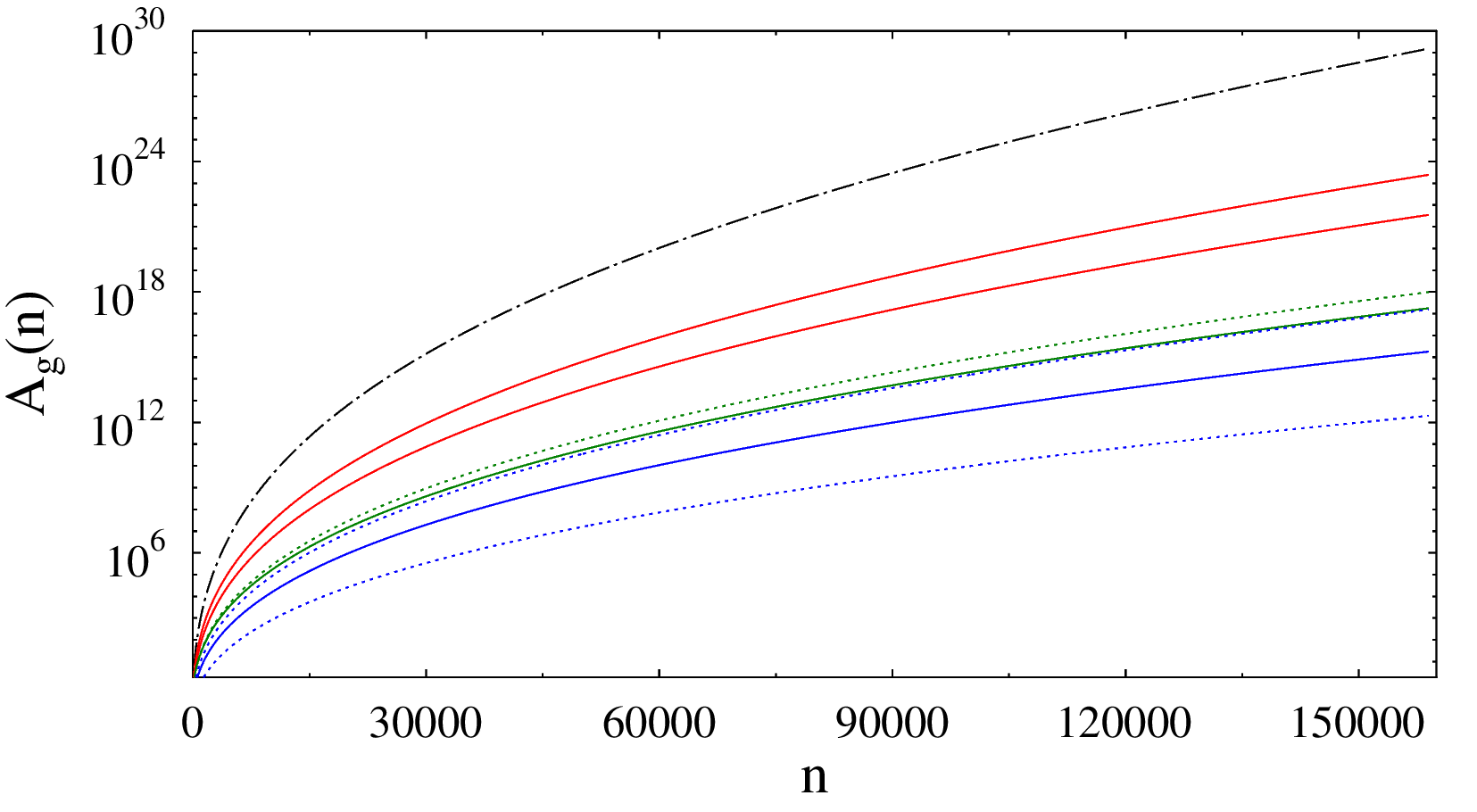}
\caption{Semi-classical amplitudes $A_g(n)$ on a logarithmic scale (see text for
details).}
\label{fig:amp}
\end{figure}

The relative weights of the generations can be understood from Fig.\ \ref{fig:amp}, where we plot the amplitudes $A_g(n)$ on a logarithmic scale. The long-dashed top line gives the amplitude of generation zero, which is identical with $\overline Q^{(2)}(n)$. The solid (s) and short-dashed (s-d) lines give, from top to bottom, the amplitudes of the generations 2 (s), 1 (s), 6+7 (s-d), 3+4+5 (s), 8 (s-d), 10 (s), and 9 (s-d). The  amplitudes of the generations 3 and higher are seen to be smaller than those of generations 1 and 2 by 2 - 3 orders of magnitudes for $n\gtrapprox5000$. These in turn are smaller than $\overline Q^{2)}(n)$ by 2 - 3 orders of magnitude, demonstrating the relative smallness of the oscillating part.

\begin{figure}[h]
\centering
\includegraphics[width=14cm,angle=0]{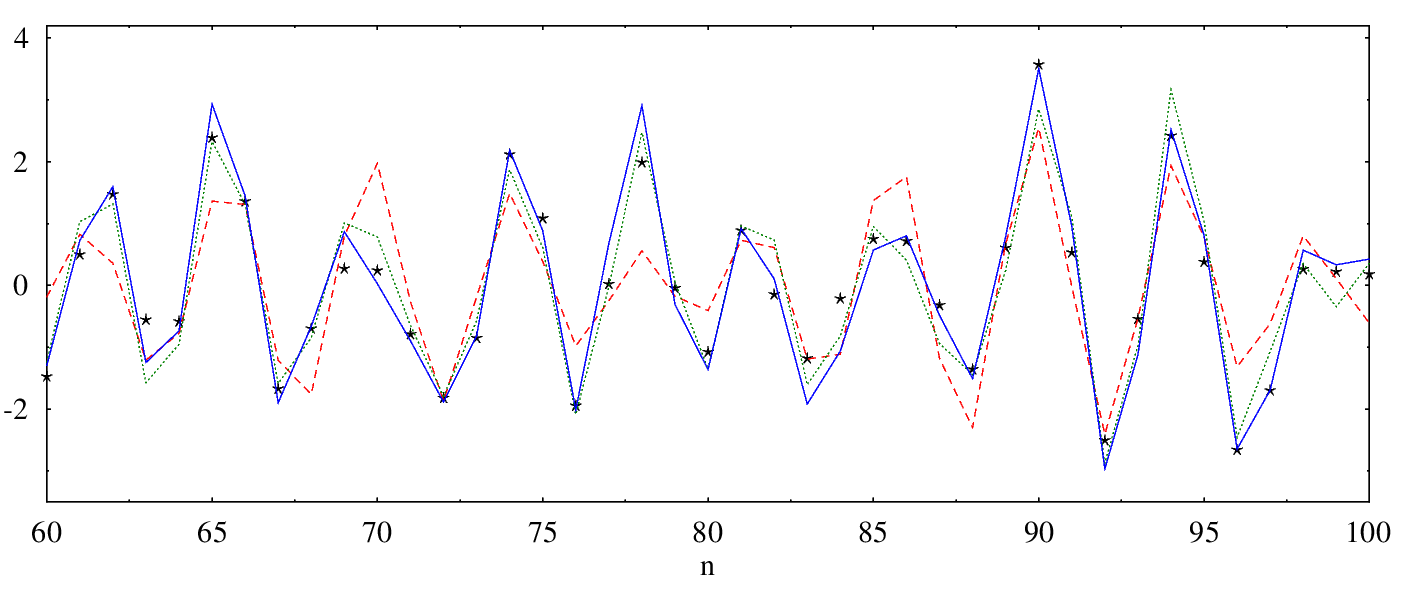}
\put(-430,90){$\delta Q(n)$}
\caption{Result of trace formula (\ref{eq:dgsc}) around $n=80$ for increasing numbers of included          generations. Dashed line (green): generations 1 and 2; dotted line (blue): generations 1-         solid line (red): generations 1-10. The stars (black) show the exact $\delta Q(n)$.}
\label{fig:conv}
\end{figure}

The relative importance of the higher orbits can be seen in Fig.\ \ref{fig:conv}, where we show the result around $n=80$. With an increasing number of generations, the amplitude of the semi-classical $\delta Q(n)$ increases slightly, but even here the differences are small. Together, these two figures demonstrate the overall rapid convergence of the trace formula.

We conclude that the oscillations in $\delta Q(n)$ are completely governed by the orbits of generations  1 and 2 with frequencies 4 and 5, which are members of the PT (3,4,5). The period of the rapid oscillations is roughly that of the orbit with the largest amplitude$^\dagger$ (i.e., $\tau_2=\pi/2$ with frequency 4), while the beat comes from the difference in their periods: the period $\Delta n = 20$ of the beat is but the inverse difference of their periods: $1/4-1/5=1/20$.\footnote{The superposition of two cos functions: $a_1\cos(x_1)+a_2\cos(x_2)$ with different periods and similar but not equal amplitudes $a_1>a_2$ yields a beat structure where the rapid oscillation is governed by the period of the component with the larger amplitude ($a_1$). Only if $a_1=a_2$, the rapid oscillation has the average period of the two components.} 
The contributions from all higher generations are practically negligible for $n\gtrapprox 4000$ and still very small around $n\sim 1000$.

\begin{figure}[h]
\centering
\includegraphics[width=10cm,angle=0]{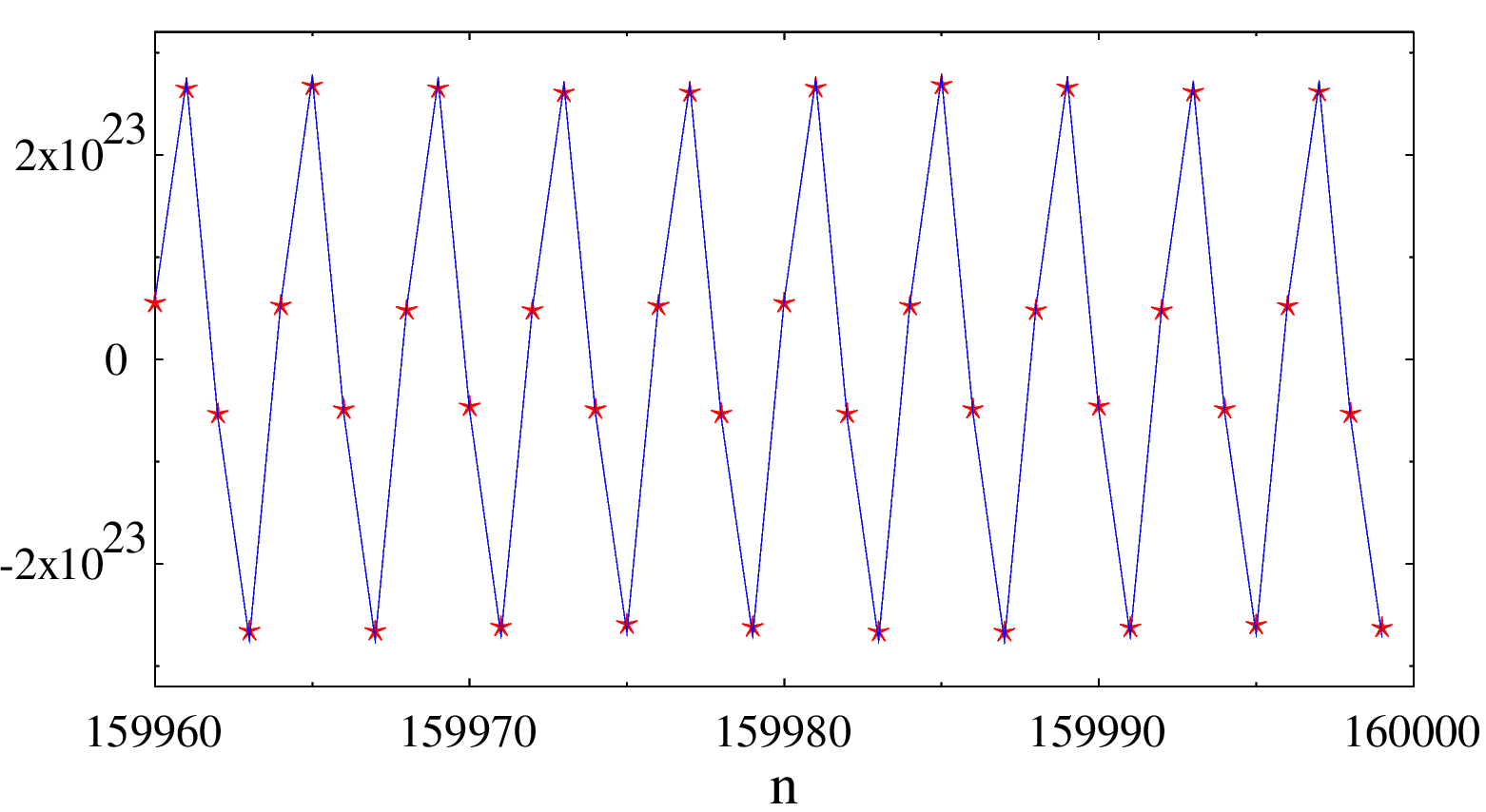}
\put(-300,90){$\delta Q(n)$}
\caption{Result of Eq.\ (\ref{eq:dgsc}), shown by the dashed (blue) line, using only the pair of orbits          of generation 2 (frequency 4), in the region near $n=160,000$. The exact $\delta Q(n)$ are shown by the (red) stars. Note that the beat structure in the exact $\delta Q(n)$ has practically disappeared.}
\label{fig:hi}
\end{figure}

For $n\gtrapprox 100,000$, the beat structure fades away and the oscillations are practically given by the pair of orbits of generation 2 (frequency 4) alone, as shown in Fig.\ \ref{fig:hi} for $n\sim 160,000$. The exact values $\delta D(n)$, shown the by stars, exhibit practically no more beating amplitude. This is due to the fact that the amplitude of generation 1 (frequency 5) here is nearly 2 orders of magnitude smaller than that of generation 2 (frequency 4). From this result we can give the rate of disappearance of the relative oscillations as
\begin{equation}
\left|\frac{\delta Q(n)}{\overline Q^{(2)}(n)}\right| \sim A_4/A_0 
\sim 2\sqrt{\kappa_0/\kappa_2}\,e^{-3(\Lambda_0-\Lambda_2)\,n^{1/3}}
\sim 1.9187\,e^{-0.25866\,n^{1/3}}.
\quad \hbox{ for } n \rightarrow \infty\,.
\label{eq:disapp}
\end{equation}
In the region $n\gtrapprox 100,000$, the truncated series of three SP corrections in (\ref{eq:rhof2cor}) does not converge fast enough; this shows up in the numerical results by a slight asymmetry of the numerical $\delta Q(n)$ with respect to its average which ought to be zero. We found that a renormalized value ${\tilde c}_3=0.025$ of the coefficient $c_3$ given in Eq.\ (\ref{eq:rhof2cor}) makes up for this asymmetry, yielding a correct average value $\rho^{(2)}_{as}(n)$ for all $n$ up to the limit $n=160,000$ of our data base for the exact $Q_2(n)$. Of course all the oscillations vanish as $n\rightarrow \infty$ as it should be expected. 

In summary, we have shown that a semiclassical trace formula, based on the periodic orbit theory, is capable of reproducing the oscillations in the distinct square partitions almost quantitatively. In particular, in the asymptotic limit of large $n$, it becomes almost exact demonstrating that the semiclassical methods developed in physics may be successfully applied to the partitions in number theory. To our knowledge, this is also the first time that a quantitative disappearance rate, as given in Eq.\ (\ref{eq:disapp}), of asymptotic oscillations of a partition, as seen in Fig.\ \ref{fig:hi}, and their exact period have been determined. 

\newpage
\section{Asymptotic Prime Partitions}
\label{sec:prime}

In section \ref{sec:int-part}, we have established that the asymptotic quantum density of states $\rho(E)$ is the same as the asymptotic number partition $P(n)$ when $E=n$. This was demonstrated with examples and comparing the asymptotic and exact results. It is therefore obvious that the methods of statistical mechanics may be applied to any partition of a positive integer $n$. In this section we discuss the partitions of $n$  into {\it primes $p$}. We first discuss the unrestricted prime partitions and then the distinct prime partitions.   Some results of the unrestricted prime partitions have been known and some are new, as outlined in Ref.\ \cite{bib:bbbm}.  A complete formula with prefactors for the asymptotic distinct prime partitions was, as far as we are aware, obtained for the first time in Ref.\ \cite{bib:mbbb}. Here we review  briefly the results of both these cases.

The many-body system that we start with is one in which the non-degenerate single-particle levels are simply given by primes $p=2,3,5,\ldots$ in dimensionless units. The total energy of the many particle system $E=n$ is then a sum of primes. The corresponding density of states is the number of partitions $P(n)$. As outlined in section \ref{sec:mpds}, to apply the method of statistical mechanics, we convert this sum into a continuous integral using the known density of primes which is deduced from the well-known prime number theorem \cite{bib:primetheorem}. We do not know if any physical system exists whose Hamiltonian has a single-particle spectrum given by primes, unlike the two cases discussed in Sec.\ \ref{sec:int-part}. There have been attempts to construct Hamiltonians whose eigenvalues are primes. Unfortunately, their potentials keep changing upon inclusion of more primes and have a fractal-like character \cite{bib:primespectra} (for detailed reviews, see \cite{bib:berry, bib:bohigas}). For now we leave this problem aside and focus on partitions of integers into primes. The following discussion mainly follows the account given in \cite{bib:bbbm} (see also \cite{bib:OEIS-u}).

\subsection{Unrestricted prime partitions}

To set the notation and to outline the method, we begin with a discrete single-particle spectrum $\{\epsilon_i;~i=1,2,3,\cdots,\}$,  where all the $\epsilon_i$ are primes.  The total energy is given by $E=\sum_{i=1}^{\infty} n_i\epsilon_i$, where $n_i$ (as before) is the occupancy  of the level $\epsilon_i$.  In general for any given single-particle spectrum, there are many ways of partitioning the energy $E$ into summands of $\epsilon_i$ or, equivalently, of partitioning positive integers into sums of primes. We begin here with unrestricted prime partitions, i.e., $n_i$ can take any positive  integer value allowed by the total energy $E$.

The canonical partition function in the $N\rightarrow \infty$ limit  is 
\begin{equation}
Z(\beta) = \prod_{{p}} \frac{1}{[1-e^{-\beta p}]}\,,
\label{eq:zofp}
\end{equation}
where the product runs over all primes $p$. This is also the generating function of the prime partitions. Taking the logarithm of (\ref{eq:zofp}) gives a sum over $p$ which we may rewrite as an integral 
\begin{equation}
\ln Z(\beta) = -\sum_p \ln (1-e^{-\beta p}) = -\int_{x_0}^{\infty} dx\,g(x) \ln(
1-e^{-\beta x})\,,
\label{eq:zex}
\end{equation}
where $g(x)$ is the exact density of primes; hereby $x_0$ is any real number smaller than the lowest prime: $x_0<p_1$=2. 

\subsection{Average prime density}

\begin{figure}[h]
    \centering
    \includegraphics[width=0.9\linewidth]{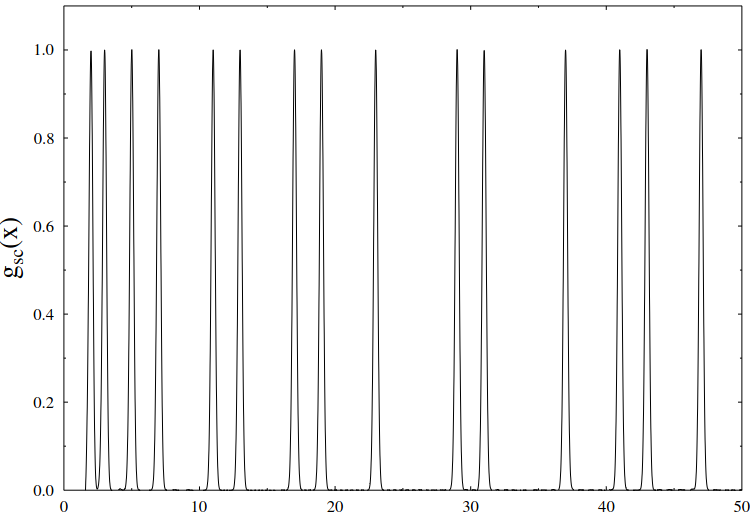}
    \caption{Density of primes $g(x)$ obtained by the semiclassical expression $g_{sc}(x)$ in Eq.\ (\ref{eq:gxsc}), using the lowest 3000 Riemann zeros $\alpha$ and maximum m= 14, coarse-grained with a Gaussian width $\gamma$ = 0.1.}
    \label{fig:gsc}
\end{figure}

The density of primes\footnote{The unpublished results of this section were done in collaboration with Ken-ichiro Arita.} is related to the prime counting function $\pi(x)$ by 
\begin{equation}
    g(x)=\frac{d\pi(x)}{dx}.
    \label{eq:gofp}
\end{equation}
An expression for $\pi(x)$ which was conjectured by Riemann in 1859 \cite{bib:riemann} and proved by  Mangold in 1959 is the following (see \cite{bib:edwards}):
\begin{equation}
\pi(x)=\sum_{m=1}^\infty \frac{\mu(m)}{m}\,J(x^{1/m})\,,  
\label{eq:pip}
\end{equation}
where $\mu(m)$ is the Moebius function [with $\mu(1)=1$], and the function $J(x)$ is given by \cite{bib:edwards}
\begin{equation}
J(x) = \sum_{n=1}^\infty \frac{1}{n}\,\sum_{p} \,\Theta(x-p^n) \,,  \qquad (x>0)
\label{eq:Jx}
\end{equation}
and $\Theta(x)$ is the standard step function [$\Theta(x)=1$ for $x\geq 0$, $\Theta(x)=0$ for $x<0$]. In (\ref{eq:Jx}), the sum over $p$ runs over all primes and $n$ is summed over all integers. Using an expansion  of $J(x)$ derived by Riemann, one obtains the following expression for the density of primes (cf.\ \cite{bib:berry})
\begin{equation}
g_{sc}(x) = \frac{1}{x\ln x} \sum_{m=1}^\infty \frac{\mu(m)}{m}
              \left[\,x^{1/m}-\frac{1}{(x^{2/m}-1)}
              -2\,x^{1/2m}\sum_\alpha \cos\left(\frac{\alpha}{m}\ln x\right)\right].
\label{eq:gxsc}
\end{equation}
Here $\alpha>0$ are the zeros of the Riemann zeta function along the positive half-axis, and the validity of the Riemann hypothesis has been assumed. This expression, which does not appear to be widely known, has the form of a semiclassical trace formula \cite{bib:book,bib:gutz}. Ideally, Eq.\ (\ref{eq:gxsc}) should yield the exact prime density $g(x)$ if the sum over $\alpha$ is not truncated and if the Riemann hypothesis is true. We have tested Eq.\ (\ref{eq:gxsc}) numerically in order to convince ourselves of its validity.\footnote{There may be a hint to a possible Hamiltonian in the structure of the density of primes derived by Riemann and further elaborated by Edwards \cite{bib:edwards}. The prime spectrum can be reproduced from the non-trivial zeros of the Riemann $\zeta$ function. As remarked in section \ref{sec:semiclass}, semiclassical trace formulae can give insights into the connection between a quantum spectrum and the periodic orbits of the corresponding classical Hamiltonian.}   Figure \ref{fig:gsc} shows the results, obtained using the lowest 3000 Riemann zeros $\alpha$. We see that the coarse-grained trace formula indeed reproduces the Gaussian-smoothed density of primes, replacing the delta functions by Gaussians centred exactly at the primes p. (Note that the sum over m can be truncated for any finite value of x; in the situation described here, maximum  m = 14 was sufficient.)

For the study of asymptotics, we replace the exact $g(x)$ by the average prime density $g_{av}(x)$ which should suffice to obtain the leading contributions. As a specific choice, we use for $g_{av}(x)$ the asymptotic prime density which is the leading term of (\ref{eq:gxsc}). 
\begin{equation} 
g_{av}(x) = 1/\!\ln(x)\,.
\label{eq:gavofx}
\end{equation} 
By substituting this function for $g(x)$ in (\ref{eq:zex}), we define the logarithm of the average partition function
\begin{equation}
\ln Z_{av}(\beta) = -\int_{a}^{\infty} dx\,g_{av}(x) \ln(1-e^{-\beta x})\,,
\label{eq:zav}
\end{equation}
where the constant $a$ must be chosen carefully. This is sufficient to obtain asymptotic results, as will be demonstrated later. 

The integrand of (\ref{eq:zav}) has a pole at $x=1$ which becomes relevant when $a<1$. We therefore define the following principal-value integral
\begin{equation}
Z_{av}(a,\beta) = - \lim_{\epsilon\to 0} \left[ \int_a^{1-\epsilon} dx\, \frac{1}{\ln
(x)}\ln(1-e^{-\beta x})
                               + \int_{1+\epsilon}^\infty dx\, \frac{1}{\ln(x)}\
ln(1-e^{-\beta x}) \right], 
                               \quad (a \neq 1)  
\label{eq:Iab}
\end{equation}

This integral exists for any $a\neq 1$ and for finite $\beta$. Details of the evaluation of the principal-value integral are given in \cite{bib:bbbm}. Here we just reproduce the result. The asymptotic form of the logarithm of the partition function is given by
\begin{equation}
\ln Z_{as}(\beta) = \frac{1}{\beta\ln(\beta)}\int_{0}^{\infty} dy\,
                    \left(1+\frac{\ln(y)}{\ln(\beta)}\right)\ln(1-e^{-y})
                  = -\frac{f_1}{\beta\ln(\beta)}+\frac{f_2}{\beta\ln^2(\beta)}\,
\label{eq:Zas}
\end{equation}
with
\begin{eqnarray}
f_1 &=& \frac{\pi^2}{6}, \nonumber \\f_2 &=& \frac{C\pi^2}{6}+\sum_k\frac{\ln(k)}{k^2} =
\frac{\pi^2}{6}[12\ln(\!A)-\ln(2\pi)]=1.887029965\dots,\,, 
\label{eq:f1f2boson}
\end{eqnarray}
where $C=0.577216$ is the Euler constant and $A= 1.282427129100$... is the Glaisher-Kinkelin constant (see OEIS-A074962 in \cite{bib:OEIS-gk}).  Note that the result (\ref{eq:Zas}) does not depend on the precise value of $a$. We have numerically tested this in Ref.\ \cite{bib:bbbm}. As a consequence, the most general form of the integral may be written in the form 
\begin{equation}
\ln Z_{as}(\beta) = \frac{1}{\beta}\sum_{k=1}^\infty 
(-1)^k\frac{f_k}{\ln^k(\beta)}\, ,
\label{eq:Zasg}
\end{equation}
where $f_k$ are simply real numbers. Eq.\ (\ref{eq:Zas}) is an excellent asymptotic approximation to the exact $\ln Z(\beta)$ in the small-$\beta$ limit.

Using the analytical approximate form (\ref{eq:Zas}) of the partition function, the inverse Laplace transform of the level density can now be evaluated in the saddle-point approximation as outlined at the end of Sect.\ \ref{sec:mpds}. 

The procedure adopted in \cite{bib:bbbm} is reproduced in the appendix \ref{appendix2}. Following which we get the asymptotic prime partition with corrections to the leading order result as 
\begin{equation}
{P}_{as}(n) = \frac{1}{2[3\ln(n)]^{1/4}n^{3/4}}\,
                   \exp \left\{ 2\pi\sqrt{\frac{n}{3\ln(n)}} 
                   \left[1-\frac{1}{2}\frac{\ln[\ln(n)]}{\ln(n)}
                   +\frac{f_2/f_1+\ln(\pi/\sqrt{3})}{\ln(n)}\right] \right\}.  
\label{eq:partas-1}
\end{equation}
It is important to point out here that the sub-leading term $ln[ln(n)]/ln(n)$ also appears in the result obtained by Vaughan\cite{bib:vaughan} but with coefficient $+1$ instead of $-1/2$ here. It has been shown that indeed there was an error in his calculation and the latter is the correct coefficient.  We test the quality of the saddle-point approximation numerically in what follows. To compare with the exact partition, we have to generate a data base of exact prime partitions.

\subsection{Numerical study of asymptotic partitions}
We generate the data base of exact prime partition ${P}(n)$ using a well-known method. Given an integer $n$, find the distinct primes that divides $n$. The sum of distinct prime factors that decompose n is denoted by $\mathscr{S}(n)$ \cite{bib:oeis-pf}. For example, $\mathscr{S}(4) = 2$ since $4 = 2 \cdot 2$ has only one distinct prime that divides it; $\mathscr{S}(6) = 5$ since $6 = 2 \cdot 3$, or $\mathscr{S}(52) = 15$ since $52 = 2 \cdot 2 \cdot 13$. (Note: if a prime factor occurs several times, it should only be counted once.) Once the sum of prime factors $\mathscr{S}(n)$ is generated in a table, the following recursion relation \cite{bib:SP95} is used to compute the prime partitions (without any restriction)
\begin{equation}
  {P}(n) = \frac{1}{n} \left[\mathscr{S}(n) + \sum_{k=1}^{n-1} \mathscr{S}(k) 
          \cdot {P}(n-k)\right].
\label{eq:pexact}
\end{equation}
which involves all prime partitions of integers less than $n$. This procedure is very time consuming for large $n$. We have been able to compute ${P}(n)$ for $n$ up to $8'654'775$. But, as we shall see, even that large number is not sufficient to reach the asymptotics of ${P}(n)$.

Using the data base for the exact ${P}(n)$ derived as above, we now test various approximations for their asymptotic behaviour. Rather than calculating the exponentially growing full function ${P}(n)$, we look at its logarithm. We compare numerically the logarithm of the exact ${P}(n)$ with that of the following approximations:

\begin{itemize}

\item To lowest order (LO), we set the prefactor of the exponent in (\ref{eq:partas-1}) to unity, ignoring its denominator, and just keep the leading exponential term
\begin{equation}
{P}_0(n) = \exp\left\{2\pi\sqrt{\frac{n}{3\ln(n)}}\right\},
\label{eq:lnp0}
\end{equation}
an asymptotic result that has been known for a long time \cite{bib:roth,bib:yang}.

\item Next we investigate the full asymptotic result (\ref{eq:partas-1}) and compare with the LO result as well as with the approximation given by Vaughan\cite{bib:vaughan} where the coefficient of the sub-leading term  is $+1$ instead of $-1/2$ as remarked earlier. 
\end{itemize}

\begin{figure}[h]
\centering
\includegraphics[angle=0,width=0.8\columnwidth,clip=true]{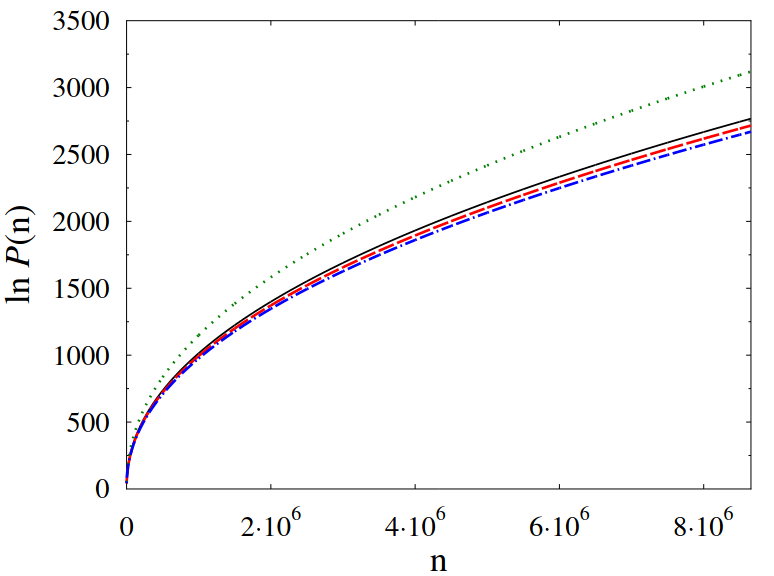}
\caption{Logarithms $\ln {P}(n)$ in various approximations. Solid line (black): exact numerical values. Dashed line (red): $\ln {P}_{as}(n)$ (\ref{eq:partas-1}), dash-dotted line (blue): LO $\ln {P}_0(n)$ (\ref{eq:lnp0}). The result of Ref.\ \cite{bib:vaughan} shown by the dotted (green) line, exhibits an increasing divergence with exact values.}  
\label{fig:p1} 
\end{figure}

We first plot $\ln {P}(n)$ versus $n$ for the various approximations in Fig.\ \ref{fig:p1}. The solid (black) curve gives the exact values $\ln {P}(n)$. Our full asymptotic approximation (\ref{eq:partas-1}), shown by the dashed (red) line, comes closest to it, improving over the lowest-order approximation $\ln {P}_0(n)$ (\ref{eq:lnp0}) shown by the dash-dotted (blue) line. In comparison, the approximation of Vaughan, shown by the dotted (green) curve, overshoots the exact values substantially. 

\begin{figure}[h]
\centering
\includegraphics[angle=0,width=.95\columnwidth,clip=true]{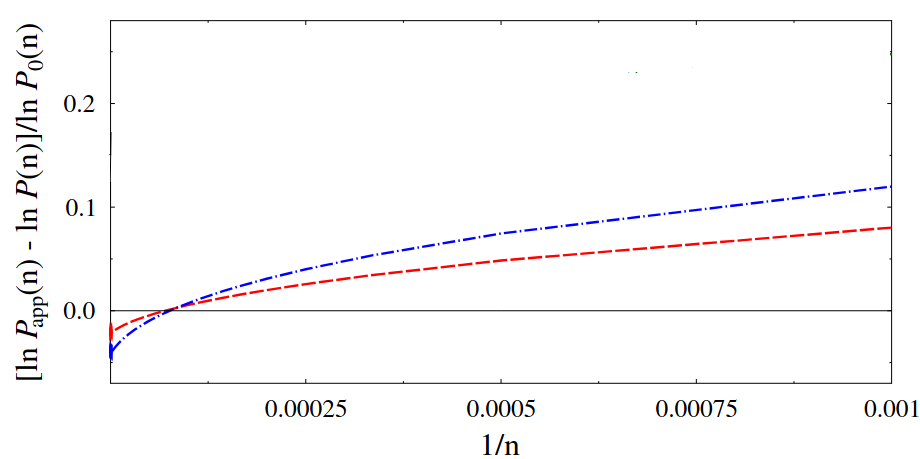}
\caption{Relative differences $[\ln {P}_{app}(n)-\ln {P}(n)]/\ln {P}_0(n)$ plotted versus $1/n$. Dashed line (red): NLO result (\ref{eq:partas-1}), dash-dotted line (blue): LO result (\ref{eq:lnp0}).} 
\label{fig:p2} 
\end{figure}

However, we are unable to assess from this figure the way in which the various approximations approach the correct asymptotic partition. To this purpose we show in Fig.\ \ref{fig:p2} the relative differences of the approximated logarithms, $[\ln {P}_{app}(n)-\ln {P}(n)]/\ln {P}_0(n)$, and plot them versus $1/n$ so that they should tend to zero for $n\to \infty$. Shown are, with the same symbols (and colours) as in Fig.\ \ref{fig:p1}, the approximation (\ref{eq:partas-1}) with NLO terms, and the LO result (\ref{eq:lnp0}). 
 
\begin{figure}[h]
\centering
\includegraphics[angle=0,width=1\columnwidth,clip=true]{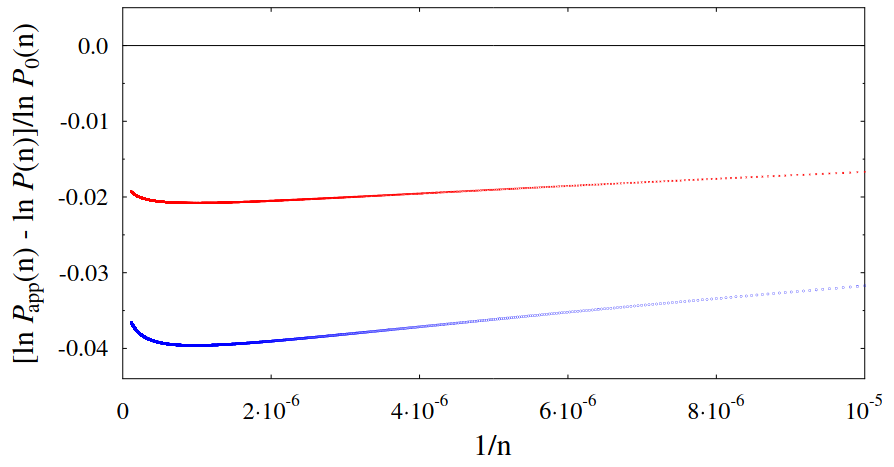}
\caption{Same as in Fig.\ \ref{fig:p2} in the lowest region $1/n \leq 10^{-5}$. (Vaughan's result does not appear at this scale.)} 
\label{fig:p4} 
\end{figure}

On this scale we still cannot see what goes on for larger $n$, but we notice sign changes occurring in the two lowest curves: at $n\sim 5'800$ for (\ref{eq:partas-1}), and at $n\sim 13'000$ for (\ref{eq:lnp0}). In order to see how (or if) the two curves approach the asymptotic value 0, we zoom into the largest region of $n$ available from our computations and further reduce the scale to $1/n \leq 10^{-5}$, finding the results seen in Fig.\ \ref{fig:p4}. 

Clearly, the differences are still not reaching zero, even for our largest value $n_{max}=8'654'775$, although the curves do bend up towards zero for $1/n\to 0$. We must therefore ask how far one has to go for the logarithm of our theoretically well-founded asymptotic result (\ref{eq:partas-1}) to go over into the exact $\ln {P}(n)$. The slopes of the curves at their ends (corresponding to $n_{max}$) are still rather small, so that there may be a very long way to go -- too long perhaps to be covered by any numerical computation of the exact ${P}(n)$.

We conclude that our result ${P}_{as}(n)$ in (\ref{eq:partas-1}) appears to have the correct asymptotic behaviour, but that even the included corrections beyond the LO are not sufficient to reach the exact partitions in the numerically accessible region.

\paragraph{Summary of the results:}

The main results of this section may be summarised as follows:

\begin{itemize}

\item While the main asymptotic form Eq.\ (\ref{eq:lnp0}) has been known for some time, we have derived non-leading order (NLO) corrections to the exponent. There has not been much discussion in the literature of the prefactor to the exponential form (\ref{eq:partlo}) needed to calculate the absolute value of the prime partitions. An exception is Vaughan \cite{bib:vaughan} who derived the prefactor and also a correction to the exponent in (\ref{eq:lnp0}). We obtain the same prefactor but must correct his NLO contribution to the exponent as given in Eq.\ (\ref{eq:partas-1}). This brings a considerable improvement for the asymptotic compared to that of Vaughan. We have also obtained analytically a the next-order correction to the NLO result \cite{bib:bbbm} which, to the best of our knowledge, is not found in the literature.

\item A well-known algorithm has been used to compute the exact prime partitions in order to compare analytical expressions for asymptotic prime partitions numerically. We have been able to do this up to more than $n\sim$ 8 millions. We believe that a numerical comparison of exact results and asymptotics has not been done up to this range before \cite{bib:bbbm}.

\item As seen earlier in Sec.\ \ref{sec:int-part}, the correction to the exponent falls off exponentially, rendering the asymptotic density almost exactly. However, as seen from Eq.\ (\ref{eq:partas-1}), the correction in the case of primes falls off logarithmically, which explains why the asymptotic limit is reached more slowly for prime partitions as compared to integer partitions $P(n)$.
 
\item Although both the exact ${P}(n)$ and the asymptotic form ${P}_{as}(n)$ given in (\ref{eq:partas-1}) are monotonously increasing, the difference is not monotonic. In fact, we found that ${P}_{as}(n)$ crosse ${P}(n)$ around $n\sim 5'800$ and approaches it from below for $n\to\infty$ (within the limits of our data). 

\item Our main conclusion is that our result ${P}_{as}(n)$ given in (\ref{eq:partas}) appear to have the correct asymptotic behaviour, but that even the corrections included beyond the LO expression ${P}_0$ in (\ref{eq:lnp0}) are not sufficient to reach the exact ${P}(n)$ in the numerically accessible region. This is left as a problem for the future. 

\end{itemize}


\subsection{Distinct prime partitions}

We now consider distinct prime partitions \cite{bib:mbb} for which not many results are known. Only  the form  of the exponential \cite{bib:roth} is known; the prefactor and corrections to LO are not known as far as we are aware. 

As in the case of bosons or unrestricted case, consider a large number $N$ of ``fermions" whose single-particle spectrum is given by the primes $p$. As in earlier cases the $E=\sum_{p} n_p\, p,$ where the energy is expresses in some dimensionless units. Here and in the following, the sums $\sum_p$ run over all primes $p$, and $n_p$ are the fermionic occupancies of the levels which must be zero or one, such that $\sum_p n_p = N\,, n_p=0,1$ The number of possible energy partitions of $E$ is denoted by $Q_N(E)$.  In particular we are interested in the limit of $N\rightarrow \infty$ and in this limit $Q(n)$ denotes the {\it distinct prime partition}.

The {\it quantum-statistical partition function} $Z^F(\beta)$ is in this limit given by
\begin{equation}
Z^F(\beta)=\prod_p [1+e^{-\beta\,p}]
\label{eq:zfinfty}
\end{equation}
where $\beta=1/kT$ is the inverse temperature and the product runs over all  primes $p$ as before. Taylor expanding the exponential in (\ref{eq:zfinfty}) and reordering the terms yields the alternative form of the partition function
\begin{equation}
Z^F(\beta)=\sum_{n=0}^\infty Q(n)\,e^{-n\beta}\,,
\label{eq:zgen}
\end{equation}
which in number theory is known as the {\it generating function} of  $Q(n)$. In the On-line Encyclopedia of Integer Sequences (OEIS) \cite{bib:OEIS-d}, the sequence of numbers $Q(n)$ is called the sequence A000586. Its first ten members are $Q(n)$ = 1, 0, 1, 1, 0, 2, 0, 2, 1, 1 for $n=0,...,9$, where $Q(0)=1$ by definition. Note also that the $Q(n)$ are a subset of the  unrestricted (bosonic) prime partitions $P(n)$, called the sequence A000607 in \cite{bib:OEIS-u}.

\paragraph{Asymptotic partition function from saddle-point approximation:}

The procedure is identical to the unrestricted case with the partition function given by Eq.\ (\ref{eq:zfinfty}). We first rewrite the inverse Laplace transform (\ref{eq:zfinfty}) by taking the natural log of $Z^F$ into the exponent:
\begin{equation}
\rho^F(E) = \frac{1}{2\pi i}\int_{-i\infty}^{+i\infty} d\beta\,
            e^{\beta E + \ln Z^F(\beta)}\,.
\label{eq:invlap}
\end{equation}
We now evaluate this integral using the saddle-point method as in the unrestricted case. The canonical entropy function is given by
\begin{equation}
S^F(E,\beta)=\beta E + \ln Z^F(\beta)\,.
\label{eq:sdef}
\end{equation}
Applying the saddle-point method to (\ref{eq:invlap}) requires to find a stationary point $\beta_0$ of the function $S^F(E,\beta)$ by solving the {\it saddle-point equation} given by 
\begin{equation}
\left. \frac{\partial S^F(E,\beta)}{\partial \beta}\right|_{\beta_0}
     = E+\frac{\partial Z^{F}(E,\beta_0)/\partial\beta}{Z^F(E,\beta_0)}=0\,.
\label{eq:spc}
\end{equation}
If this equation has a solution $\beta_0$, which will be a function $\beta_0(E)$, one evaluates the successive partial derivatives of $S^F(E,\beta)$ at $\beta_0$:
\begin{equation}
S^{F}_n(E,\beta_0) = \left. \frac{\partial^n S^F(E,\beta)}{\partial \beta^n}
                    \right|_{\beta_0}.
\end{equation} 

The natural log of the partition function given in Eq.\ (\ref{eq:zfinfty})
\begin{equation}
\ln Z^F(\beta) = \sum_{p=2}^{\infty}\ln\left(1+e^{-\beta p}\right)\, , 
\label{eq:lnzbex}
\end{equation}
which is approximated by the integral 
\begin{equation}  
\ln Z^F(\beta) \sim \int_{2}^{\infty}dx\, g_{av}(x) \ln\left(1+e^{-\beta x }\right)\, , 
\label{eq:lnzb}
\end{equation}
where $g_{av}(x)=\frac{1}{\ln(x)}$ is the approximate density of primes obtained using prime number theorem given in Eq.\ (\ref{eq:gavofx}). If the exact density $g(x)$ is used then the integral would give the exact result (\ref{eq:lnzbex}). The evaluation the integral in the limit $\beta\to 0$ follows closely the method outlined in \cite{bib:bbbm}. 

It is straightforward to evaluate the integrals analytically, and we obtain
\begin{equation}  
\ln Z^F_{as}(\beta) = \frac{1}{\beta\ln(\beta)}\left[-\frac{\pi^2}{12}
                     +\frac{1}{\ln\beta}\left(\frac{C\pi^2}{12}
                     +\sum_k(-1)^{k-1}\frac{\ln(k)}{k^2}\right)\right]\!,
\label{eq:lnzas}
\end{equation}
where $C=0.5772156649\!\dots$ is the Euler constant. This expression has a similar form to that of Eq.\ (\ref{eq:Zas}) as in the case of unrestricted partitions but with different numerical coefficients.

\paragraph{Solution of saddle-point equation and $Q_{as}(n)$}
\label{sec:speq}

In order to find the saddle point $\beta_0$, we start from  the entropy $S^F(\beta)$ in the asymptotic limit. Using Eqs.\ (\ref{eq:sdef}) and (\ref{eq:lnzas}) we get up to order $1/(\ln\beta)^2$
\begin{equation}  
S^F(E,\beta) = \beta E-\frac{F_1}{\beta\ln(\beta)}
               +\frac{F_2}{\beta\,(\ln\beta)^2}+\cdots
\label{eq:sfbeta}
\end{equation}
where 
\begin{equation}  
F_1 = \frac{\pi^2}{12}\,, \qquad 
F_2=\left[\frac{C\pi^2}{12}+\sum_{k=1}^\infty (-1)^{k-1}
                                 \frac{\ln(k)}{k^2}\right]
\end{equation}
The infinite sum in $F_2$ may be expressed in a closed form in terms of  a derivative of the Riemann zeta function, leading to
\begin{equation}
F_2=\frac{\pi^2}{12}[C-\ln(2)]-\frac{\zeta'(2)}{2}=0.3734242774\!\dots
\end{equation}
Eq.\ (\ref{eq:sfbeta}) is identical in form with that of the bosonic case given in Eq.\ (\ref{eq:slo}) obtained in Ref.\cite{bib:bbbm} :
\begin{equation}  
S^B(E,\beta)= \beta E-\frac{f_1}{\beta\ln(\beta)}
            + \frac{f_2}{\beta(\ln\beta)^2}+\cdots\,,
\label{eq:sbbeta}
\end{equation}
where $f_1,f_2$ are explicitly given in Eq.\ (\ref{eq:f1f2boson}).

The only difference in going from bosonic to the fermionic case is that the coefficients $f_1$ and $f_2$ of \cite {bib:bbbm} are replaced here by the $F_1$ and $F_2$, respectively. Therefore we obtain our  result simply by replacing the coefficients $f_i$ in the bosonic case by the $F_i$ in the present fermionic case and following the steps outlined in \cite{bib:bbbm} and explicitly given in Appendix B. 

Thus we can directly give the result for the fermionic case as
\begin{equation}
\overline\rho^F(E) = \frac{1}{{\sqrt{4E^{3/2}[6\ln(E)]^{1/2}}}}\,
                \exp\left\{ 2\pi\sqrt{\frac{E}{6\ln(E)}}
                \left[1-\frac{1}{2}\frac{\ln[\ln(E)]}{\ln(E)}
                +b^F\frac{1}{\ln(E)}\right]\right\}
\label{eq:rhofe}                
\end{equation}
with the constant
\begin{equation}
b^F = \left[\frac{F_2}{F_1}+\ln(\pi/\sqrt{6})\right]=0.7028796287\!\dots
\end{equation}
The asymptotic $Q_{as}(n)$ is then obtained replacing $E$ by $n$ above, so that:

\begin{equation}
Q_{as}(n)=\frac{1}{{\sqrt{4n^{3/2}[6\ln(n)]^{1/2}}}}\,
                \exp\left\{ 2\pi\sqrt{\frac{n}{6\ln(n)}}
                \left[1-\frac{1}{2}\frac{\ln[\ln(n)]}{\ln(n)}
                +b^F\frac{1}{\ln(n)}\right]\right\}.
\label{eq:qasn} 
\end{equation}
The corresponding result for the bosonic partitions is given in Eq.\ (\ref{eq:partas-1}) taken from \cite{bib:bbbm}. Note that the leading exponential terms and the denominators of the pre-exponential terms in (\ref{eq:qasn}) and (\ref{eq:partas}) differ by a factor $1/\!\sqrt{2}$. Furthermore, since the $Q(n)$ are a subset of the $P(n)$, their values must be smaller, which asymptotically is brought about by the extra factor $1/\!\sqrt{2}$ in the leading exponential term. The first correction term in the exponent, namely $-\frac{1}{2}\ln[\ln(n)]/\!\ln(n)$, is identical in both cases. As far as we know, the above result (\ref{eq:qasn}) for the distinct prime partitions can only be found in Ref.\cite{bib:mbb}.

Next we compare numerically the asymptotic result (\ref{eq:qasn}) with the exact values $Q(n)$ of the distinct prime partitions.

\paragraph{Numerical test of $Q_{as}(n)$}
\label{sec:num}

\begin{figure}[h]
\centering
\vspace*{-0.3cm}
\includegraphics[width=12.cm,angle=0]{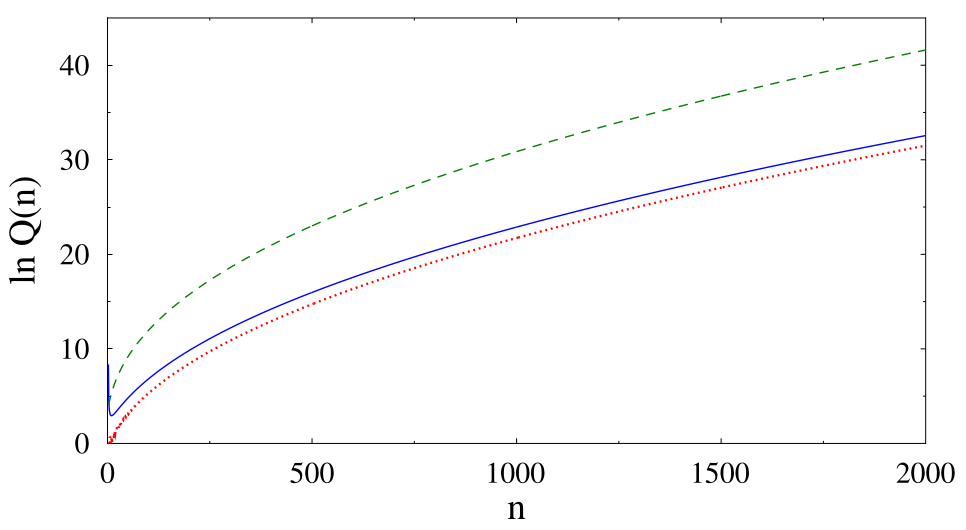}
\vspace*{-0.7cm}
\caption{Exact $\ln Q(n)$ by dots (red), lowest-order asymptotic form $\ln Q_0(n)$ from (\ref{eq:qas0}) by the dashed line (green) and the full asymptotic form $\ln Q_{as}(n)$ given by (\ref{eq:qasn}) by the solid (blue) line, shown as functions of $n$ up to $n=2000$.}
\label{fig:qofnlow}
\end{figure}

\begin{figure}[h]
\centering
\vspace*{-0.3cm}
\includegraphics[width=12.cm,angle=0]{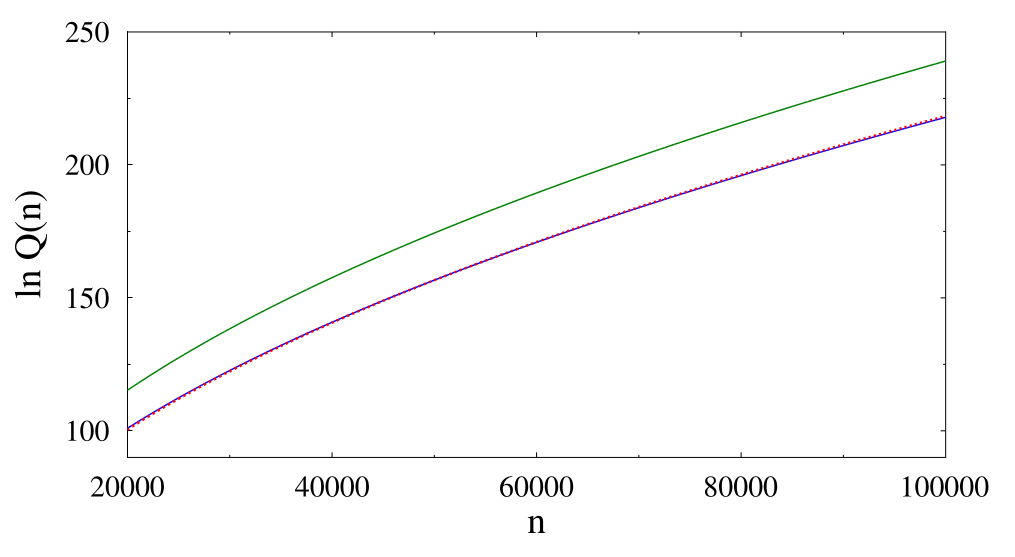}
\vspace*{-0.7cm}
\caption{Same as Fig.\ \ref{fig:qofnlow} in the region $n=20\,000-100\,000$.}
\label{fig:qofn}
\end{figure}

Here, we test the  asymptotic result (\ref{eq:qasn}) numerically. We have generated the exact $Q(n)$ up to $n=100\,000$. In Figs.\ \ref{fig:qofnlow} and \ref{fig:qofn} we show their values by the dots (red) on a logarithmic scale in two regions of $n$. The dashed line (green) shows the leading-order exponential expression  
\begin{equation}
Q_0(n)=\exp\left\{2\pi\sqrt{\frac{n}{6\ln(n)}}\right\}\,,
\label{eq:qas0} 
\end{equation}
which is the only asymptotic result that has been given so far in the literature\cite{bib:roth}, and the solid (blue) line gives our full asymptotic result (\ref{eq:qasn}). 

\noindent
A large discrepancy between $Q_0(n)$ and $Q(n)$ is noticed for all $n$. The full asymptotic result $Q_{as}(n)$ (\ref{eq:qasn}) approaches the exact $Q(n)$ much better (except in the academic limit $n\to 0$ where it diverges due to the pre-exponential factor). In Fig.\ \ref{fig:qofn} for the values $n \geq 20\,000$, the two curves can hardly be distinguished. 

A considerable improvement over the simple exponential form (\ref{eq:qas0}) is achieved by (\ref{eq:qasn}). A closer look reveals that the curve for $Q_{as}(n)$, which for smaller $n$ overestimates the exact $Q(n)$ and crosses the latter at $n=50000$. As noticed in the previous section, a similar result was found in \cite{bib:bbbm} for the bosonic prime partitions, where $P_{as}(n)$ crosses  $P(n)$ much earlier and then tends to approach it asymptotically from below for $n\to\infty$. 

\begin{figure}[H]
\centering
\vspace*{-0.3cm}
\includegraphics[width=12.cm,angle=0]{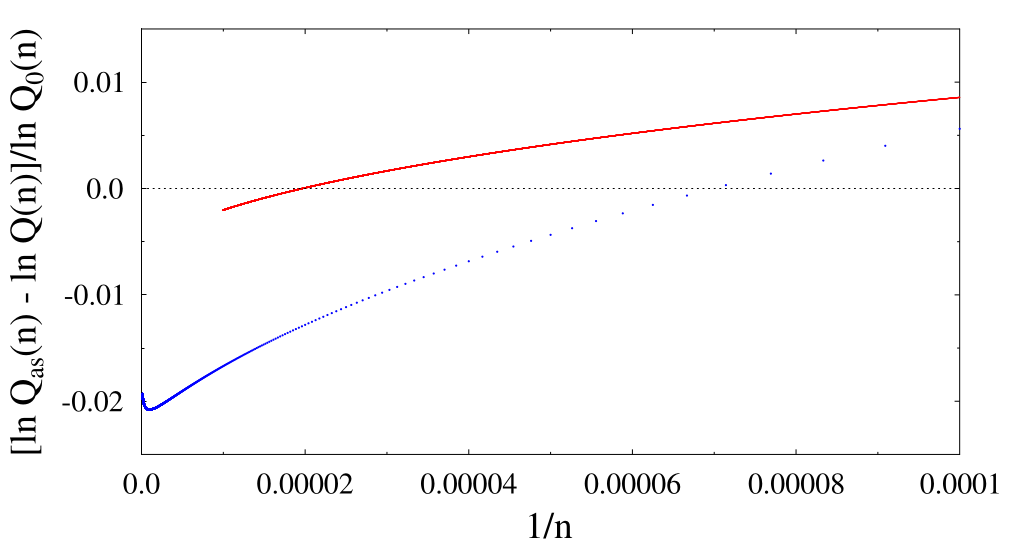}
\vspace*{-0.2cm}
\caption{Relative difference $[\ln Q_{as}(n)-\ln Q(n)]/\ln Q_0(n)$, shown versus $1/n$ by the solid (red) line. The dotted (blue) line shows the corresponding quantity obtained in \cite{bib:bbbm} for the bosonic prime partitions $P(n)$.}
\label{fig:qperr}
\end{figure}

In order to focus on this asymptotic behaviour, we show in Fig.\ \ref{fig:qperr} the difference of the natural  logs relative to the lowest-order term, i.e., the quantity $[\ln Q_{as}(n)-\ln Q(n)]/\ln Q_0(n)$, plotted  versus $1/n$ in a region of the largest $n$ available. The solid (red) curve gives the result obtained with the full asymptotic form (\ref{eq:qasn}). For comparison we show in this figure by the dotted (blue) curve shown in Fig.\ \ref{fig:p4} quantity obtained in the unrestricted (bosonic) prime partitions $P(n)$ and their respective asymptotic forms. The overall behaviour of the two curves is similar. For the results in \cite{bib:bbbm} we had larger values of $n$ available. There we noticed a tendency for the difference to approach zero from below for $1/n\to 0$ (i.e.\ $n\to\infty$), as can be recognized from the blue curve in Fig.\ \ref{fig:qperr}.

While we could not show the approach to zero since we had not gone high enough in $n$, after the announcement of our results on the arXiv in 2019 \cite{bib:mbbb}, V. Kotesovec performed numerical studies of our $\ln Q_{as}(n)$ and the exact $\ln Q(n)$, computing these quantities up to $n_{max}=10^8$. In order to do this, he programmed in the assembler language a special floating point arithmetic in which both the mantissa and the exponent of these quantities have 8 bytes. By this procedure it was possible to generate 10 million terms in 9 hours, and the calculation of $n_{max}=10^8$ terms took 31 days. V.\ Kotesovec kindly sent us his results which confirm our findings (red line) up to $n=10^5$ and furthermore show that, indeed, the difference $[\ln Q_{as}(n)-\ln Q(n)]/\!\ln Q_0(n)$ tends towards zero from the same side as that of the bosonic (unrestricted) prime partitions (blue curve). Similarly to the situation for the latter, the asymptotic ratio $\ln Q(n)/\!\ln Q_{as}(n)$ first exceeds the value 1 but then reaches a maximum, occurring here at $n = 14\,474\,250$, has an inflection point at $n \gtrapprox 33\,272\;000$, and gradually decreases back towards 1. One or two million terms are far from enough for this finding; it is necessary to have at least 40 million terms. A graph of Kotesovec's result for $\ln Q(n)/\!\ln Q_{as}(n)$ is posted at OEIS (see \cite{bib:mbb}).

In summary, we have shown how an improved asymptotic expression for the function $Q(n)$, which counts the number of distinct prime partitions of an integer $n$, can be obtained from asymptotic expansions of the partition function $Z^F(\beta)$ in (\ref{eq:zgen}) and the corresponding density of states $\rho^F(E)$ in (\ref{eq:rhofe}). The asymptotic $Q_{as}(n)$ overestimates the exact $Q(n)$ for smaller $n$ but overshoots it for $n \gtrapprox 50,000$. Like in \cite{bib:bbbm}, the limit $Q_{as}(n)\to Q(n)$ for $n\to\infty$ cannot be demonstrated rigorously. However, forcing the calculation of $Q_{as}(n)$ and of $Q(n)$ up to $n=10^8$, V.\ Kotesovec \cite{bib:kotes} has shown numerically that $Q_{as}(n)$, indeed, approaches $Q(n)$ monotonously for $n \gtrapprox 4\times 10^7$ \cite{bib:kotes} assuming that the difference can be approximated by a term $c_2/\ln^2(n)$ in the square brackets of the asymptotic expansion (\ref{eq:qasn}) and showed that the results are very sensitive to the value of $c_2$. We join his suggestion that the systematic evaluation of the algebraic value of $c_2$, or of other correction terms in (\ref{eq:qasn}), could be a topic of interesting future research for the next generation of patient researchers.

But we can state that already with our present result (\ref{eq:qasn}), we have obtained an excellent asymptotic approximation for the distinct prime partitions which is much better than the hitherto only known result \cite{bib:roth}.


\section{Summary}
\label{sec:summary}
We are now celebrating not only the centenary of the foundations of Quantum Mechanics, but also of Quantum  Statistical Mechanics (QSM) whose roots go back to the paper by Bose in 1924. Hardy and Ramanujan explored the  {\it asymptotic partitions of integers of various kind} a few years before the birth of quantum statistical mechanics. Bethe derived the formula for the {\it density of states in the levels of heavy nucleus} a decade later. These seemingly disparate developments were shown to be related by Auluck and Kothari in their work on {\it Statistical mechanics and partition of numbers}.This review, largely based on the work done earlier by the authors and their collaborators, explores these conversations between quantum statistical mechanics and theory of integer partitions further while also recording some new and hitherto unknown results.

The review begins with a discussion of the partition function of a many body system in QSM and its relation to the quantum density of states which encodes how the total energy of the system is distributed among particles of the many body system. In general energy is a real number. However, when the energy takes only integer values the quantum density of states gets related to partitions of integers of various kind. As a result the methods of statistical mechanics may be used to obtain results in number partitions. While this can yield in principle exact results, in practice the Laplace inversion is difficult if not impossible. We approximate the inversion using the saddle point approximation which yields leading asymptotic result for the density. A method for getting the corrections order by order in the asymptotic expansion  is discussed to complete the basic framework. 

Within this framework, we discuss various types of partitions. We reproduce the well known results for unrestricted integer partitions as well as distinct partitions drawing parallels with the many body density of states of particles obeying Bose and Fermi statistics. The generalisation of this is  the Gentile statistics where the occupancy varies from distinct to unrestricted; the density of states of Gentile statistics nicely interpolates from Bose to Fermi occupancies with analogous partitions when the energy is an integer. Interestingly, the distinct square partitions of integers has remarkable structure with regular oscillations and beats. This is unlike other partitions where there are variations around the mean but nothing that shows the kind of regularity in the distinct square partitions. We investigate the origin of such regular behaviour and find that its origin is in frequencies dominated by the Pythagorean triplets. We then provide an approximate trace formula which reproduces such a regular behaviour. 

Finally we have presented the partitions of integers into primes, both unrestricted as well as distinct partitions. We present results which give excellent asymptotic approximations to the exact results in both cases, going beyond the scantily known results. 

\medskip

\paragraph{Acknowledgement:} The series of results presented in this review would not have possible without the fundamental ideas of R.K. Bhaduri who is no more with us, having passed away in 2019 when the distinct prime partitions were still being investigated. This review is written to honour him. We thank Muoi Tran, C. S. Srivatsan and Johann Bartel for their collaboration in the works is  presented here. We thank Shouvik Sur and Rajesh Ravindran for their assistance in the initial stages of work on prime partitions. We are grateful to Florin Spinu who, in private communication, confirmed our NLO contribution in unrestricted prime partitions and also located the error in Ref.\cite{bib:vaughan}. We are grateful to Ken-ichiro Arita for his collaboration on the semiclassical derivation of the density of primes found in section 7.2.  Finally, we acknowledge stimulating correspondence with  V. Kotesovec and, in particular, the communication of his most recent numerical results.  Our Thanks to C.S. Sundar for careful reading and corrections. 

We are grateful to the IMSc (Chennai, India), the McMaster University (Hamilton, Canada)and the University of Regensburg (Regensburg, Germany) for their hospitality at different times during the course of this research.

\newpage

\appendix 

\section{Gentile statistics and coloured partitions}
\label{appendix1}

In what follows we  generalise the analysis presented in Sect.\ref{sec:int-part} to arbitrary spectra but with a single quantum number with occupancies as defined in Gentile statistics.  As before the particles confined to a potential whose single particle spectrum is a monomial of the form $m^s\,, ~~m\ge 0$\cite{bib:muoi}.   The spectral information is sufficient for further analysis even if the corresponding physical system is unknown. Suppose the excitation energy $E=n$. We may distribute this energy among particles in many different ways, a particular distribution may be written as
\begin{equation}
    n=m_1^s+m_2^s+\cdots m_k^s\,=\,\sum_{i=1}^{k} m_i^s \,,
    \label{eq:n-sum-k}
\end{equation}
where $m_i$ in the summand are all positive integers since the spectrum has only positive energy levels and excitation energy is always deemed to be positive. Such a summand always exists but there are restrictions on the minimum number of terms for a given exponent $s$ \footnote{This appears in the Waring problem proposed by Edward Waring in 1770: ``each natural number $k$ has an associated positive integer $s$ such that every natural number is the sum of at most k natural numbers raised to the power s". Let $g(s)$ denotes the minimum value of $k$ for every $s$, then trivially we have g(1)=1, and furthermore g(2)=4, g(3)=9, g(4)=19, etc. as proved by different authors in the 20th century ( The sequence A002804 in the OEIS)\cite{bib:waring}.}. 

The relevant physical system for obtaining partitions is thus one in which at zero temperature all the particles are in the ground state whose energy is set to zero. At any excitation energy, particles are excited from the ground state and distributed in the single particle states with maximal occupancy given by $k$ in each state as shown in Fig.\ \ref{fig:partbosons-6}. The number of ways in which such a distribution can be achieved is the density of states at that energy (same as the number of ways of partitioning the energy).

The {\it partition function} for such a system in the asymptotic limit of the number of particles $N\rightarrow \infty$ is given by
\begin{eqnarray}
Z_{\infty}(\beta) &=&
\prod_{m=1}^{\infty}[1+\exp(-\beta m^s) + \cdots +\exp(-k\beta
m^s)]\nonumber\\
&=&~\prod_{m=1}^{\infty}~\sum_{n=0}^{k}\exp(-n\beta m^s)\, ,
\label{eq:zprod}
\end{eqnarray}
where the single particle spectrum is given by a power law. We note that the partition function in Eq.\ (\ref{eq:zprod}) is indeed the grand-canonical partition function when the chemical potential $\mu=0$. In particular we have 
\bea
Z_{\infty}(\beta) &=&~
\prod_{m=1}^{\infty}\frac{1}{[1-\exp(-\beta m^s)]}\,;~~ \mbox{unrestricted}  \nonumber\\
Z_\infty(\beta)&=&~\prod_{m=1}^{\infty}~[1+\exp(-\beta m^s)]\,; ~~\mbox{distinct}~~ k=1
\eea
corresponding to the bosons (unrestricted) and fermions (distinct)\cite{bib:muoi} discussed in Sect.\ref{sec:int-part}. Note that with $k$ arbitrary this is indeed the partition function for the well-studied Gentile statistics\cite{bib:gentile}\footnote{Note that while many forms of quantum statistics have been studied, asymptotically free particles obey only Bose or Fermi statistics. The extensions are useful in discussing nature of quasi-particle excitations as in the case of anyons and Haldane exclusion statistics\cite{bib:anyons}}

Setting $x=\exp(-\beta)$, the partition function may be written as
\begin{equation}
Z_{\infty}(x)=\sum_{n=1}^{\infty} P_k^{(s)}(n) x^n
=\prod_{m=1}^{\infty}\frac{[1-x^{(k+1)m^s}]}{[1-x^{m^s}]}.
\label{eq:sis}
\end{equation}
In the asymptotic limit of large number of particles, the degeneracy $P_k^{(s)}(n)$ is the number of ways of partitioning $n$.  The partition function is then the generating function of partitions $P_k^{(s)}(n,k)$. Using Eq.\ (\ref{eq:entropy}), we obtain
\begin{equation}
S=\beta E+\sum_{m=1}^{\infty}~\ln[1-\exp(-(k+1)\beta
m^s)]-\sum_{m=1}^{\infty}~\ln[1-\exp(-\beta m^s)].
\label{eq:sinfty}
\end{equation}
We now evaluate the sums approximately using the Euler-Maclaurin series. We have
\begin{eqnarray}
&-&\sum_{m=1}^{\infty} \ln[1-\exp(-\beta m^s)]=-\sum_{m=0}^{\infty}
\ln[1-\exp(-\beta (m+1)^s)]\nonumber\\
&=&
-\int_{0}^{\infty} \ln[1-\exp(-\beta (y+1)^s)]dy
-\frac{1}{2}\ln[1-\exp(-\beta)]
+s\sum_{k=1}^{\infty} \frac{B_{2k}}{2k(2k-1)},
\label{eq:em}
\end{eqnarray}
where $B_{2k}$ are the Bernoulli numbers. The integral in the Euler-Maclaurin expansion may be evaluated as follows:
\begin{eqnarray}
&-&\int_{0}^{\infty} \ln[1-\exp(-\beta (y+1)^s)]dy 
=-\int_{0}^{\infty} \ln[1-\exp(-\beta y^s)]dy
+\int_{0}^{1} \ln[1-\exp(-\beta y^s)]dy \nonumber \\
&=&
\frac{C(s)}{\beta^{1/s}}
+\ln[1-\exp(-\beta)] - s\int_{0}^{1} \frac{\beta y^s}{\exp(\beta y^s)
-1}dy,
\end{eqnarray}
where
\begin{equation}
C{(s)} = \Gamma(1+\frac{1}{s})\zeta(1+\frac{1}{s})
\end{equation}
in terms of the Riemann zeta function. In the {\it high temperature limit}, that is $\beta \rightarrow 0$, we have
\begin{equation}
-\int_{0}^{\infty} \ln[1-\exp(-\beta (y+1)^s)]dy
\approx \frac{C(s)}{\beta^{1/s}}
+\ln[1-\exp(-\beta)] - s + O(\beta)
\end{equation}
The series involving Bernoulli numbers may be evaluated approximately as follows: The Euler-Maclaurin series applied to logarithm of Gamma function gives,
\begin{equation}
\ln (\Gamma(n+1)) =n\ln(n) -n +\frac{1}{2}\ln(2\pi) +\sum_{k=1}^{\infty} \frac{B_{2k}}{2k(2k-1) n^{2k-1}}
\end{equation}
We may consider this as the expansion of zero by putting $n=1$ and get
\begin{equation}
s\sum_{k=1}^{\infty} \frac{B_{2k}}{2k(2k-1)}
=s - \frac{s}{2}\ln(2\pi).
\end{equation}

Keeping terms up to $O(\beta)$, and using the above arguments the last term in Eq.\ (\ref{eq:sinfty}) may be written as
\begin{equation}
-\sum_{m=1}^{\infty}[1-\exp(-\beta m^s)]
\approx \frac{C(s)}{\beta^{1/s}}
+\frac{\ln[1-\exp(-\beta)]}{2} -\frac{s}{2}\ln(2\pi)+O(\beta)~
\label{eq:euler}
\end{equation}
The infinite sum in the second term of Eq.\ (\ref{eq:sinfty}) is also evaluated similarly by scaling $\beta$ appropriately. Note that the constant term does not get scaled.

The entropy to the leading order in $\beta$ is given by
\begin{equation}
S=\beta E + \frac{\alpha C{(s)}}{\beta^{1/s}}- {1\over
2}~\ln[1-\exp(-(k+1)\beta)] +{1\over 2}~\ln[1-\exp(-\beta)]+O(\beta)~,
\label{eq:sinftylo}
\end{equation}
where
\begin{equation}\alpha = 1-\frac{1}{(k+1)^{1\over s}}.
\label{eq:alpha}
\end{equation}
We note that the expansion in powers of $\beta$ is equivalent to taking the high-temperature limit though this limit should be employed cautiously when $\beta$ is weighted by $k$ which can in principle take large values.  The last term of $ O(\beta)$ in the RHS of Eq.\ (\ref{eq:sinfty}) causes a shift in the energy E by a constant (for example 1/24 in the s=1 case) in $S$. In the asymptotic formula for the level density for large $E$, obtained by the saddle point method, this may be ignored. Note, however, that such a shift was included by Rosenzweig\cite{bib:rosen} to improve the Bethe formula\cite{bib:bethe} for the nuclear level density. We further note that the $k\rightarrow \infty$, or the bosonic limit, is special since this limit involves precisely the sum given in Eq.\ (\ref{eq:euler}) and not the difference of two series as in Eq.\ (\ref{eq:sinfty}). As a result the $\beta$ independent term, namely $-(s/2)\ln(2\pi)$, appears only in this limit and not for any finite $k$. Indeed, as we shall see below this term is crucial in getting the prefactor correctly in the Hardy-Ramanujan Formula.

The saddle point on the real axis\footnote{Note that we evaluate the saddle point only along the real axis. In general there may be other complex saddle points. We assume, without proof, that this is the only one that gives the correct asymptotic behaviour as it is known to yield the rigorous answers for ordinary partitions.} is evaluated by taking the derivative of the entropy 
\be
S_1(\beta)=E-{1\over s}~ {\alpha C{(s)}\over{{\beta^{(1+1/s)}}}}~,
\ee
where only the leading terms in $\beta$ are retained. This is sufficient for obtaining the asymptotic result. Equating the derivative to zero, the saddle-point in $\beta$ is given by
\begin{equation}
\beta_0 =~\lambda_s~ E^{-{s\over {1+s}}},~\mbox{where}~~\lambda_s=\left({\alpha C{(s)}\over s}\right)^{{s\over {1+s}}}~.
\label{betz}
\end{equation}
The entropy at the saddle is given by
\begin{equation}
    S(E)=(s+1) \lambda_s E^{1/(s+1)} - {1\over
2}~\ln[1-\exp(-(k+1)\lambda_sE^{-s/(s+1)}] +{1\over 2}~\ln[1-\exp(-\lambda_sE^{-s/(s+1)})]+O(1/E^{s/(s+1)})~,
\end{equation}

We also need the second derivative of $S$ at the saddle point and is given by
\begin{equation}
    S_2(\beta_0)=\frac{s+1}{s} \frac{\lambda_s^{1+1/s}}{\beta_0^{2+1/s}}=\frac{s+1}{s} \frac{E^{\frac{2s+1}{s+1}}}{\lambda_s}.
\end{equation}

Substituting this in the saddle point expression for the density of
states in Eq.\ (\ref{eq:asymsp11}), the asymptotic density of states is given by
\begin{equation} 
\overline{\rho}_{k}^{(s)}(E)~\approx~ \lambda_s\sqrt{s}
\frac{\exp\left[\lambda_s
(s+1)E^{{1\over
{1+s}}}\right]}{\sqrt{2\pi(s+1)E^{(3s+1)/(s+1)}[1-\exp(-(k+1)\lambda_s
E^{-s/(s+1)})]}}~,
\label{eq:rhofgen}
\end{equation}
where the subscript $k$ in $\overline{\rho}$ indicates the main property
of the statistical system under consideration. We have kept the
exponential term under the square-root sign in the denominator to
indicate the interpolation property of the density of states between the
Bose ($k\rightarrow \infty$) and the Fermi ($k=1$) limits even though
keeping the exponential may not be consistent with the order of the
expansion. The correct expression is given below.

\begin{figure}[ht]
    \centering
    \includegraphics[width=0.9\linewidth]{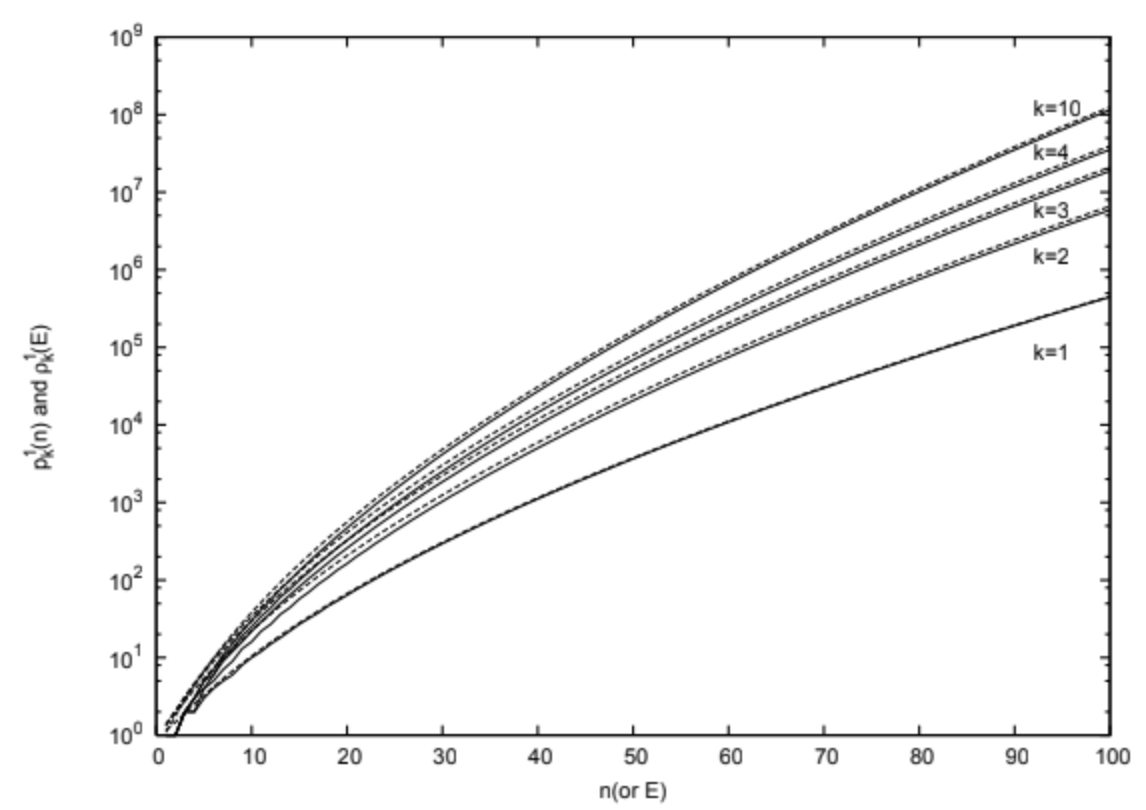}
    \caption{Comparison of the exact $P_k^{(1)}(n)$ (solid line) and the asymptotic $\overline\rho_k^{(1)}(E)$ (dashed line) for s = 1.}
    \label{fig:gentiles-1}
\end{figure}

For finite $k$ there always exists an energy $E$ which is large enough
such that the following approximation is useful:
\begin{equation} \overline{\rho}_{k}^{(s)}(E)~\approx~ \sqrt{s\lambda_s}
\frac{\exp\left[\lambda_s
(s+1)E^{{1\over
{1+s}}}\right]}{\sqrt{2\pi(s+1)(k+1)E^{(2s+1)/(s+1)}}}~,
\label{eq:rhofty1}
\end{equation}
Note that the above expression is identical to the Eq.\ (23) for $\bar\rho_k^{(s)}(n)$ 
in \cite{bib:muoi} corresponding to the special case of $k=1$. The degeneracy 
or $k$ dependence in the above equation is also hidden in the parameter 
$\lambda_s$. An asymptotic formula derived in Ref.\ \cite{bib:kutta} based 
on Gentile statistics for the special case of $s=1$ is the same as the one 
given in Eq.\ (\ref{eq:rhofty1}). However, to the best of our knowledge, no 
general formula for arbitrary $s$ and $k$ had been derived earlier. A comparison of the exact and asymptotic partitions for $s=1$ for some sample values of $k$ are shown 
in Fig. \ref{fig:gentiles-1}. 
The detailed analysis  given above in this section is meant as a template 
to obtain specific, some well known, results discussed in the main text.

\section{Prime partitions: Saddle point solution}
\label{appendix2}
In this appendix we illustrate the method used to obtain the leading asymptotic prime partitions along with non-leading order corrections. In order to find the extremum we shall isolate the most singular terms in $S(\beta)$ in the high-temperature limit. We first write entropy, using (\ref{eq:Zas}) above, in the form
\begin{equation}
S(\beta)= \beta E-\frac{f_1}{\beta\ln(\beta)}+\frac{f_2}{\beta\ln^2(\beta)},
\label{eq:slo}
\end{equation}
where we for simplicity omit the subscript ``$as$'' henceforth as it is implicit. Since the entropy above is given up to order $1/\ln^2(\beta)$,  all further calculations will be done up to this order. The first two derivatives of entropy are given by 
\begin{equation}
S_1{(\beta)}= E+\frac{f_1}{\beta^2\ln(\beta)} +\frac{f_1}{\beta^2\ln^2(\beta)}
 -\frac{f_2}{\beta^2\ln^2(\beta)}+\cdots,
\label{eq:slo1}
\end{equation}
\begin{equation}
S_2(\beta)= -\frac{2f_1}{\beta^3\ln(\beta)}-\frac{3f_1}{\beta^3\ln^2(\beta)}
+\frac{2f_2}{\beta^3\ln^2(\beta)}+\cdots. 
\label{eq:slo2}
\end{equation}
The saddle-point solution $\beta_0$ is given by the following convenient form
\begin{equation}
\beta_0 E=-\frac{f_1}{\beta_0\ln(\beta_0)}
+\frac{f_2}{\beta_0\ln^2(\beta_0)}
-\frac{f_1}{\beta_0\ln^2(\beta_0)}
+\cdots. 
\label{eq:spcond}
\end{equation}
This is a transcendental equation whose solution can be obtained iteratively as outlined below. However, we may use the above condition directly in $S(\beta_0)$ to obtain
\begin{equation}
S(\beta_0)= 2\beta_0 E+\frac{f_1}{\beta_0\ln^2(\beta_0)}+\cdots,
\end{equation}
and 
\begin{equation}
S_2(\beta_0)= \frac{1}{\beta_0^2}\left[2\beta_0 E
-\frac{f_1}{\beta_0\ln^2(\beta_0)}+\cdots\right]. 
\end{equation}
Using the above solutions in terms of as yet undetermined $\beta_0$, we obtain
the asymptotic density is  given by
\begin{equation}
\overline\rho(E) = \frac{\exp(2\beta_0 E+\frac{f_1}{\beta_0\ln^2(\beta_0)}+\cdots)}{\sqrt{(2\pi/\beta_0^2)\left[2\beta_0 E-\frac{f_1}{\beta_0\ln^2(\beta_0)}+\cdots\right]}}
\label{eq:rhoprime}
\end{equation}
This asymptotic density is the same as the asymptotic formula for prime partitions denoted by ${P}(n)=\overline\rho(n$=$E)$. Any further analysis requires the solution of the saddle-point condition $S_1(\beta)=0$ for $\beta_0(E)$. Since this is a transcendental equation, its solution is not straight forward, even with the approximations we have made. We present the main results below leaving the details to Ref.\cite{bib:bbbm}.

\paragraph{Solution to leading order(LO)}: 

The solution of the saddle-point equation, keeping only the leading-order
\begin{equation}
\beta_0 E=-\frac{f_1}{\beta_0\ln\beta}\, ~~~\mbox{where}~~~ f_1=\frac{\pi^2}{6}\, ,
\label{eq:spcondlo}
\end{equation}
is given by\cite{bib:vaughan, bib:bbbm}
\begin{equation}
\frac{1}{\beta_0}=
\sqrt{\frac{3}{\pi^2} E\ln(E)}\,.
\label{eq:solnlo}
\end{equation}
To leading order, therefore, we have the following result for the density, or equivalently for unrestricted prime partitions:
\begin{equation}
\overline\rho(E)=\frac{e^{S(\beta_0) }}{\sqrt{2\pi S_2(\beta_0)}}
=\frac{e^{2\pi\sqrt{E/[3\ln(E)]}}}{\sqrt{4E^{3/2}[3\ln(E)]^{1/2}}}\,.
\label{eq:partlo}
\end{equation}

The leading exponential form has been  well known for quite some time,  that is  $\ln[\rho(E)]\approx 2\pi\sqrt{E/(3\ln E)}$ \cite{bib:roth, bib:yang} .  The prefactor was first given by Vaughan \cite{bib:vaughan} by calculating $\sqrt{2\pi S_2(\beta)}$ which agrees with the calculation presented in \cite{bib:bbbm}. Next we consider corrections to the the asymptotic result given in Eq.\  (\ref{eq:partlo}).

\paragraph{Non-leading order (NLO) corrections:}

The results given up to the LO may be further improved by including additional terms that were neglected in Eq.\ (\ref{eq:spcond}).  This is done by assuming the trial solution to be of the  form 
\begin{equation}
\beta_0=\pi\sqrt{\frac{1}{3E\ln(E)}}
\left[1+a\,\frac{\ln[\ln(E)]}{\ln(E)}+b\,\frac{1}{\ln(E)}\cdots\right],
\end{equation}
where $a,b$ are undetermined coefficients to be determined using the equation above up to order $1/ln(E)$. The form of the solution is suggested by the transcendental equation (\ref{eq:spcond}) itself. Since the LHS of (\ref{eq:spcond}) is a monomial in $E$, the only way this can be satisfied is to have additional corrections to cancel the non-leading terms. The saddle point condition to go beyond the LO is given in Eq.\ (\ref{eq:spcond}) which is rewritten in a more convenient form given by
\begin{equation}
E=-\frac{f_1}{\beta_0^2\ln\beta_0}\left[1+\frac{1-f_2/f_1}{\ln\beta_0}+O(1/\ln^2\beta_0)\right].
\label{eq:spcondnlo}
\end{equation}
The solution of this equation up to the order of $1/\ln(E)$ is given in \cite{bib:bbbm}
\begin{equation}
\beta_0
=\pi\sqrt{\frac{1}{3E\ln(E)}}
\left[1-\frac{1}{2}\,\frac{\ln[\ln(E)]}{\ln(E)}+
\frac{\ln(\pi/\sqrt{3})+(\frac{f_2}{f_1} -1)}{\ln(E)}\cdots\right].
\label{eq:solnnlo}
\end{equation}
Substituting this in Eq.\ (\ref{eq:rhoprime}), we have the solution at NLO given by 
\begin{equation}
\overline\rho(E) = \frac{\exp \left\{ {2\pi\sqrt{\frac{E}{3\ln(E)}}}
\left[1-\frac{1}{2}\frac{\ln[\ln(E)]}{\ln(E)}+
\frac{f_2/f_1+\ln(\pi/\sqrt{3})}{\ln(E)}+\cdots\right]
\right\}} {\sqrt{\{4[3\ln(E)]^{1/2}E^{3/2}+\cdots\}}}\,.
\label{eq:partnlo}
\end{equation}
Identifying $\overline\rho(E)$ with ${P}(n$=$E)$, the above equation gives the asymptotic prime partitions of an integer $n$. The asymptotic result for prime partitions of an integer $n$ is then given by
\begin{equation}
{P}_{as}(n) = \frac{1}{2[3\ln(n)]^{1/4}n^{3/4}}\,
                   \exp \left\{ 2\pi\sqrt{\frac{n}{3\ln(n)}} 
                   \left[1-\frac{1}{2}\frac{\ln[\ln(n)]}{\ln(n)}
                   +\frac{f_2/f_1+\ln(\pi/\sqrt{3})}{\ln(n)}\right] \right\}.  
\label{eq:partas}
\end{equation}
A few comments are in order here:
\begin{itemize}

\item The leading term in the exponential, namely $2\pi\sqrt{\frac{n}{3\ln(n)}}$ agrees with previously known results 
      \cite{bib:roth,bib:yang,bib:vaughan}.
\item The prefactor given by $\frac{1}{2[3\ln(n)]^{1/4}n^{3/4}}\,$ agrees with the analysis given in Ref.\ 
      \cite{bib:vaughan}.
\item The first correction to the exponent given above, proportional to $\ln[\ln(E)]/\!\ln(E)$, is similar to that given by Vaughan \cite{bib:vaughan}. However, its coefficient here is -$\frac12$, which is the correct coefficient as shown in Ref.\ \cite{bib:bbbm}. Vaughan has the coefficient $+1$ instead of the correct $-1/2$, which is crucial.
\footnote{see the discussion in mathOverflow on this topic:\\ \url{https://mathoverflow.net/questions/180858/wrong-asymptotics-of-oeis-a000607-number-of-partitions-of-an-integer-in-prime-p}}  
\end{itemize}


\begin{thebibliography}{99}
\bibitem{bib:kotes} V. Kotesovec: See the entry by Kotesovec on OEIS A000586-Number of partitions of n into distinct primes. By clicking on the entry, leads to the graph which shows the asymptotic approach of distinct partitions going beyond the results presented here.
\end{thebibliography}
\end{document}